\documentstyle[12pt]{article}
\catcode`@=11
\def\beqra{\begin{eqnarray}} \def\eeqra{\end{eqnarray}}
\def\beqast{\begin{eqnarray*}} \def\eeqast{\end{eqnarray*}}
\def\beq{\begin{equation}}      \def\eeq{\end{equation}}
\def\be{\begin{enumerate}}   \def\ee{\end{enumerate}}
\def\fo{\hbox{{1}\kern-.25em\hbox{l}}}
\def\fnote#1#2{\begingroup\def\thefootnote{#1}\footnote{#2}\addtocounter
{footnote}{-1}\endgroup}

\def\al{\alpha}

       \def\Del{\Delta}
\def\eps{\epsilon}     

\def\la{\lambda}       
\def\mn{\mu\nu}

\def\om{\omega}        \def\Om{\Omega}

       \def\pa{\partial}

\def\ch{\@startsection{section}{1}{\z@}{-3ex plus-1ex minus-.2ex}%
        {2ex plus.2ex}{\large\sc}}

\def\ce{{\cal E}}

\def\raisenot{\raise .5mm\hbox{/}}
\def\nota{\ \hbox{{$a$}\kern-.49em\hbox{/}}}
\def\notA{\hbox{{$A$}\kern-.54em\hbox{\raisenot}}}
\def\notb{\ \hbox{{$b$}\kern-.47em\hbox{/}}}
\def\notB{\ \hbox{{$B$}\kern-.60em\hbox{\raisenot}}}
\def\notc{\ \hbox{{$c$}\kern-.45em\hbox{/}}}
\def\notd{\ \hbox{{$d$}\kern-.53em\hbox{/}}}
\def\notbd{\ \hbox{{$D$}\kern-.61em\hbox{\raisenot}}} 
\def\note{\ \hbox{{$e$}\kern-.47em\hbox{/}}}
\def\notk{\ \hbox{{$k$}\kern-.51em\hbox{/}}}
\def\notp{\ \hbox{{$p$}\kern-.43em\hbox{/}}}
\def\notq{\ \hbox{{$q$}\kern-.47em\hbox{/}}}
\def\notW{\ \hbox{{$W$}\kern-.75em\hbox{\raisenot}}}
\def\notz{\ \hbox{{$Z$}\kern-.61em\hbox{\raisenot}}}
\def\notpa{\hbox{{$\partial$}\kern-.54em\hbox{\raisenot}}}

  \def\vk{\vec{k}}

\def\vx{\vec{x}}       
       
\def\vnab{\vec{\nabla}}

\def\7#1#2{\mathop{\null#2}\limits^{#1}}        
\def\5#1#2{\mathop{\null#2}\limits_{#1}}        

\def\lsim{\mathrel{\mathpalette\@versim<}}
\def\gsim{\mathrel{\mathpalette\@versim>}}
\def\Asl{\relax{\rm /\kern-.60em A}}
\def\Bsl{\relax{\rm /\kern-.56em B}}
\def\Dsl{\relax{\rm /\kern-.56em D}}
\def\pasl{\relax /\kern-.56em \pa}

\def\inbar{\vrule height1.5ex width.4pt depth0pt}
\def\IB{\relax{\rm I\kern-.18em B}}
\def\IC{\relax\leavevmode\hbox{\,$\inbar\kern-.3em{\rm C}$}}
\def\ID{\relax{\rm I\kern-.18em D}}
\def\IE{\relax{\rm I\kern-.18em E}}
\def\IF{\relax{\rm I\kern-.18em F}}
\def\IG{\relax\leavevmode\hbox{\,$\inbar\kern-.3em{\rm G}$}}
\def\IH{\relax{\rm I\kern-.18em H}}
\def\II{\relax{\rm I\kern-.18em I}}
\def\IK{\relax{\rm I\kern-.18em K}}
\def\IL{\relax{\rm I\kern-.18em L}}
\def\IM{\relax{\rm I\kern-.18em M}}
\def\IN{\relax{\rm I\kern-.18em N}}
\def\IO{\relax\leavevmode\hbox{\,$\inbar\kern-.3em{\rm O}$}}
\def\IP{\relax{\rm I\kern-.18em P}}
\def\IQ{\relax\leavevmode\hbox{\,$\inbar\kern-.3em{\rm Q}$}}
\def\IR{\relax{\rm I\kern-.18em R}}
\def\sed{\hbox{{\sf S}\kern-.4em\hbox{\sf S}}}
\def\ZZ{\relax{\sf Z\kern-.4em Z}}
\def\smIR{\hbox{{\footnotesize\rm I}\kern-.2em\hbox{\footnotesize\rm
R}}}
\def\smIO{\ \hbox{{\footnotesize\rm I}\kern-.4em\hbox
{\footnotesize\bf O}}}

\def\smIQ{\ \hbox{{\footnotesize\rm I}\kern-.5em\hbox
{\footnotesize\bf Q}}}

\def\IGa{\relax{\rm I}\kern-.18em\Gamma}
\def\IPi{\relax{\rm I}\kern-.18em\Pi}
\def\IQt{\relax\leavevmode\hbox{$\kern.3em\inbar\kern-.3em\Theta$}}
\def\IOm{\relax\hbox{$\kern3.48pt\inbar\kern1.8pt
\inbar\kern-5.28pt\Omega$}}

\def\ca#1{\relax\ifmmode
{\cal#1} \else$\cal#1$\fi}  
\def\Sf#1{\relax\ifmmode\hbox{\sf#1}\else{\sf#1}\fi}  
\def\fibby{\ifcase\@ptsize                    
                \font\tenrm=cmfib8\or         
                \font\elvrm=cmfib8 scaled\magstephalf\or        
                \font\twlrm=cmfib8 scaled\magstep1 \fi}         
\def\TeXey{\ifcase\@ptsize\or\or             
                \font\twlrm=cmr10 scaled\magstep1     
                \font\twlmi=cmmi10 scaled\magstep1    
                \font\twlit=cmti10 scaled\magstep1    
                \font\twlbf=cmbx10 scaled\magstep1\fi}  

\def\ch{\@startsection{section}{1}{\z@}{-3ex plus-1ex minus-.2ex}%
        {2ex plus.2ex}{\large\sc}}
\def\sch{\@startsection{subsection}{2}{\z@}{-1.5ex plus-1ex minus-.2ex}%
        {1pt plus.2ex}{\sc}}
\def\seceq{\@addtoreset{equation}{section}
        \def\theequation{\thesection.\arabic{equation}}}      
\def\lapp{\raisebox{-.4ex}{\rlap{$\sim$}} \raisebox{.4ex}{$<$}}
\def\gapp{\raisebox{-.4ex}{\rlap{$\sim$}} \raisebox{.4ex}{$>$}}
\def\con{\ifmmode \hbox{\bf*} \else{\bf*}\fi}   
\def\scon{\ifmmode\hbox{\footnotesize\rm\bf*}
\else{\footnotesize\rm\bf*}\fi}

\def\0#1{\relax\ifmmode\mathaccent"7017{#1}
        \else\accent23#1\relax\fi}              

\catcode`@=12
\begin{document}
\begin{tabbing}
\hskip 11.5 cm \= {SLAC-PUB-5741}\\
\hskip 1 cm \>{SCIPP-92-07}\\
\hskip 1 cm \>{SU-ITP-92-7}\\
\hskip 1 cm \>{\today}\\
\hskip 1 cm \>{(T/AS)}\\
\end{tabbing}
\vskip 2cm
\thispagestyle{empty}
\begin{center}
{\Large\bf    TOWARDS THE THEORY OF THE }
\vskip 0.4cm
{\Large\bf ELECTROWEAK
PHASE TRANSITION\fnote{*}{Research supported by the Department of
Energy under grants 7-443256-22411-3 and DE-AC03-76SF00515,
and by the National Science Foundation grant PHY-8612280.\\
E-mail: LINDE@PHYSICS.STANFORD.EDU; DINE, HUET, LEIGH@SLACVM}}
\vskip 1.2cm
{\bf Michael Dine and Robert G. Leigh}\\
Santa Cruz Institute for Particle Physics,\\
University of California, Santa Cruz, CA 95064\\
\vskip 0.4 cm
{\bf Patrick Huet}\\
Stanford Linear Accelerator Center\\
Stanford University, Stanford, CA 94309\\
\vskip 0.4 cm
{\bf Andrei Linde} \fnote{\dagger}{On leave from: Lebedev
Physical Institute, Moscow.}\\
Department of Physics, Stanford University, Stanford, CA 94305\\
\vskip 0.3 cm
{\bf Dmitri Linde}\\
Gunn High School, Palo Alto, CA 94305\\
\vskip 1  cm
{\em Submitted for Publication to Physical Review D}
\end{center}
\vfill
\pagebreak
\thispagestyle{empty}
\vfill

\ {}

\vskip 2.5cm
\begin{center}
{\large\bf ABSTRACT}
\end{center}
\vskip .5cm

We investigate various problems related to the theory of the
electroweak phase transition.
This includes determination of the nature of the phase transition,
discussion of the possible role of the higher order radiative
corrections and the theory of the formation and evolution of bubbles
of the new phase. We show, in particular, that no dangerous linear
terms in the scalar field $\phi$ appear in the expression for the
effective potential. We have
found,  that for the Higgs boson mass smaller than the
masses of W and Z  bosons, the phase transition is of the first
order. However, its strength is approximately 2/3 times less than
what follows from the one-loop approximation.

The phase transition occurs due to production and expansion of
critical bubbles. Subcritical bubbles may be important only if the
phase transition is very weakly first order. A general analytic
expression for the probability of the bubble formation is obtained,
which may be used for study of tunneling in a wide class of theories.

The bubble wall velocity depends on many factors, including the ratio
of the mean free path of the particles to the
thickness of the wall. Thin walls in the electroweak theory have a
nonrelativistic velocity, whereas thick walls may be relativistic.

Decrease of the cubic term by the factor 2/3 rules out baryogenesis
in the minimal version of the electroweak theory. Even though we
concentrate in this paper on the phase transition in this
theory, most of our
results can be applied to more general models as well,
where baryogenesis is possible.

\pagebreak
\section{Introduction}

Twenty years ago David Kirzhnits discovered that the symmetry
between weak and electromagnetic interactions should be
restored at the very early stages of the evolution of the universe
 \cite{[1]}. Symmetry breaking between weak and electromagnetic
interactions occurs when the universe cools down to a critical
temperature $T_c \sim 10^2$ GeV. His results were confirmed
by an investigation performed in 1974 by Weinberg, Dolan and
Jackiw and by Kirzhnits and Linde  \cite{[1a]}, and soon the
theory of the electroweak phase transition became one of the
well established ingredients of modern cosmology.  Surprisingly
enough, a complete theory of this phase transition is still
lacking.

In the first papers on this problem it was assumed that the phase
transition is of the second order  \cite{[1],[1a]}. Later
Kirzhnits and Linde showed  \cite{[1b]}
that in the gauge theories with many particles,
and especially with particles which are much more heavy than
the Higgs boson $\phi$, one should take into account corrections
to the high temperature approximation used in \cite{[1],[1a]}.
These corrections lead to the occurrence of cubic terms
$\sim g^3\phi^3 T$ in the expression for the effective potential
$V(\phi,T)$. As a result, at some temperature, $V$
acquires an extra minimum,
and the phase transition is first order  \cite{[1b]}.
Such phase transitions occur through the formation and subsequent
expansion of bubbles of the scalar field $\phi$ inside the
symmetric phase $\phi = 0$. A further investigation of this
question has shown that the phase transitions in grand unified
theories are always strongly first order  \cite{Linde81}. This
realization, as well as the mechanism of reheating of the universe
during the decay of the supercooled vacuum state suggested in
\cite{[1b],[11]}, played an important role in the development
of the first versions of the inflationary universe scenario
 \cite{Guth81}. (For a review of the theory of phase transitions
and inflationary cosmology see Ref.  \cite{[2]}.)

For a long time it did not seem likely that the
electroweak phase transition could have any dramatic consequences,
unless the  Higgs boson is exceptionally light. Even though the
possibility of a strong baryon number violation during the
electroweak phase transition was pointed out fifteen years ago
by Linde \cite{Tunn1} and by Dimopoulos and Susskind \cite{DS},
only after the groundbreaking paper by Kuzmin, Rubakov and
Shaposhnikov \cite{KRS} was it realized that such processes do
actually occur and may erase all previously generated baryon
asymmetry of the universe.

Recently, the possibility that electroweak interactions may not only
erase but also produce the cosmic baryon asymmetry has led to
renewed interest in
the electroweak phase transition. A number of scenarios have been
proposed for generating the asymmetry  \cite{5} -- \cite{12}.
All of them require that the phase transition should be
strongly first order since otherwise the baryon asymmetry generated
during the phase transition subsequently disappears. In all of these
scenarios the asymmetry is produced near the walls of the bubbles of
the scalar field $\phi$.
Thus, an understanding of the nature of the phase transition in the
electroweak theory and investigation of the structure of the bubbles
produced during the phase transition are of some importance.
For this purpose one should make a much more thorough analysis
of the electroweak phase transition than the analysis which is
necessary for an approximate calculation of the critical temperature.

Unfortunately, despite the fact that one is dealing with a weakly
coupled theory, many aspects of the phase transition are surprisingly
complicated. Indeed, the literature contains contradictory claims and
statements on almost every important question.
In this paper, we will attempt to resolve a number of these
questions, or at least to delineate the issues which are
crucial to a complete analysis.  We will confine our attention
to weakly coupled theories, not because strongly coupled
theories (e.g. technicolour theories) are not of interest, but
simply because we will find these problems quite difficult even
in theories with explicit Higgs particles. Even more specifically, we
will discuss the simplest version of the electroweak theory
containing only one Higgs boson. We doubt that the baryon asymmetry
can be generated in this simple model; in fact, we will find some new
evidence against it. However, this model will help us to illustrate
various possibilities which may be realized in more complicated
theories.

The first problem to be studied is whether the phase transition
is first order, and precisely how strongly first order it may
be.  One clear requirement arises if no net $B-L$ is produced
at the transition (in the model of  \cite{10}
a net $B-L$ is produced).  In this circumstance, as first stressed
in  \cite{5}, the rate of baryon number violating transitions after
the phase transition is completed must be smaller than the expansion
rate. In practice, this means that the ratio of the
Higgs field $\phi$ inside the bubble to the temperature $T$ cannot be
smaller than one, in order that the sphaleron energy not be too
small.

This condition was used in  \cite{5,[8]} to impose a strong
constraint on the Higgs mass in the minimal version of the
electroweak theory, $m_H  \, \,  \lapp \,  \,  42$ GeV.  This, of
course, already contradicts the present experimental limits
$m_H  \, \, \,  \gapp \,\, \, 57$
GeV  \cite{LEP}.  However,  more careful consideration of
various theoretical uncertainties indicated that the constraint
might be somewhat weaker, permitting $m_H$ up to $55$ GeV,
or possibly
higher  \cite{[4]}. In any case, successful baryogenesis almost
certainly requires some extension of the standard model, possibly
including more scalar fields.  In multi-Higgs
models  \cite{[8],8}, the limits are substantially weaker.  Indeed,
Anderson and Hall  \cite{[5]} have noted that
simply adding a scalar singlet to the model significantly weakens
the constraint.

Before one can discuss details of the process of baryogenesis,
it is necessary to check that the results of our investigation of the
phase transition are reliable. This is not a trivial issue even in
the minimal electroweak theory. Indeed, as stressed in
Refs.  \cite{[1b],[11]}, each new order of perturbation
theory at finite temperature may bring a new factor of
$g^2 T/m \sim gT/\phi$ for the
theories with gauge boson masses $m \sim g\phi$.
This means that the results of the
one loop calculations may become unreliable at $\phi  \, \,  \lapp \,
\, gT$. A
rather unexpected (and often overlooked) consequence of this
observation is that we cannot even say in
an absolutely  reliable way
that symmetry at high temperatures is completely restored; one
may say only that the strength of the symmetry breaking is
limited by the constraint $\phi < gT$, $m < g^2 T$  \cite{[2]}.
In theories with $g \ll 1$ this result is quite informative.
However,  gauge coupling constants in the electroweak theory
are not {\it much} smaller than one. Therefore, the reliability of
our results concerning the region $\phi  \, \,  \lapp \,  \,  T$
deserves a detailed investigation.

This issue became more urgent with the recent claim by Brahm and Hsu
that higher order corrections lead to the appearance of a
term in the effective potential $\sim - g^3\phi T^3$
 \cite{Hsu}. This term is linear in $\phi$; it is very large at
small $\phi$, and it removes the local minimum of
$V(\phi,T)$ created by the cubic term $\sim - g^3\phi^3 T$.
As a result, the phase transition ceases to be first order. In fact,
the phase transition ceases to be a phase transition, since the
scalar field does not vanish at any temperature. At large
temperature the scalar field remains smaller than $g T$,
which means that this result is not in direct contradiction
with our earlier conclusions. But still, the result of Ref.
 \cite{Hsu} seems somewhat surprising.

The importance of understanding these issues has
been highlighted even more dramatically by the recent
work of Shaposhnikov  \cite{Shap}.
He has also found a linear term in the effective potential,
but with an opposite sign! Shaposhnikov concludes
that the phase transition is much more strongly first
order than expected. He argues that  baryogenesis in
the minimal electroweak model is possible despite the
problems with the  CP violation and obtains an improved
constraint on the Higgs boson mass, $m_H \,\, \, \lapp \, \, \, 64$
GeV, which is quite consistent with the experimental constraints.
Thus, without a proper study of the higher order
corrections to the effective potential, one may be
unable to make any conclusions concerning the
possibility of baryogenesis in the standard model.

The authors of Refs.  \cite{Hsu,Shap} obtained linear terms
  by simply substituting the
effective masses found at one loop back into the one loop
calculation of $V(\phi,T)$.  Such a procedure is generally reliable
when calculating Green's functions, or tadpoles corresponding to
${dV\over d\phi}(\phi,T)$. However,  it leads
to combinatoric errors when calculating the free energy.
Our investigation of this problem shows that if one is
careful with counting of Feynman diagrams and with
gauge invariance, neither positive nor negative linear
terms $\sim g^3 \phi T^3$ appear in the effective potential
  \cite{OurPaper}.

Moreover,   we will argue that, despite all uncertainties
with higher order corrections, the expectation value of the
scalar field $\phi$ actually disappears, $\phi = 0$,
at a  temperature higher than some critical temperature
$T_o$. (Note, that this would be impossible in the presence of
linear terms of either sign.)

However,
 high order corrections do lead to a definite and significant
modification of the one-loop results. Namely, they lead to a
decrease of the cubic term $g^3 \phi^3 T$ by a factor $2/3$.
This effect decreases the ratio $\phi/T$ at the point
of the phase transition by approximately the same factor $2/3$.
This makes baryogenesis virtually impossible in  the context of
the minimal standard model with $m_H > 57$ GeV.

Assuming that one knows the shape of the effective potential at small
$\phi$, one should still work hard to determine the ratio $\phi/T$ at
the point of the phase transition. One needs to know at what
temperature the transition actually occurs, and some details of
{\it how} it occurs.
At very high temperatures the effective potential of the Higgs
field, $V(\phi,T)$, has a unique minimum at the symmetric
point $\phi = 0$. As the temperature is lowered, a second minimum
appears. At a critical value $T_c$, this second minimum becomes
degenerate with the first one. However, the phase transition
actually occurs at a somewhat lower temperature, due to the
formation of bubbles of true vacuum which grow and fill the universe.
The usual way to study bubble formation is to use the euclidean
approach to tunneling at a finite temperature  \cite{[3]}. One should
find high-temperature  solutions, which describe the so-called
critical bubbles. Then one should calculate their action, which leads
to an exponential suppression of the probability of bubble formation.
Typically, these calculations are rather complicated, and
analytic results can only be obtained in a few cases.  One of
these is the thin wall approximation, which is valid (as in
the case of transitions at zero temperature) if the difference in
depth of the two minima of $V(\phi,T)$ is much smaller than the
height
of the barrier between them. In this case the radius of the bubble at
the moment of its formation is much larger than the size of the
bubble
wall, and the properties of the bubble can be obtained very easily.
Recently Anderson and Hall performed a thorough analytic study of the
phase transition in the electroweak theory
in one loop approximation \cite{[5]}. Their results
for the one loop effective potential
$V(\phi,T)$ completely agree with the results of our investigation
 \cite{[4]}. However, in their study of the bubble
formation they assumed that the thin-wall approximation is
applicable. As we will see, for the Higgs boson masses in the
range of interest, $m_H \ < \  m_W$, this is not the case
 even if one takes into account modification of the
cubic terms by higher order corrections.
Fortunately, we were able to obtain a simple analytic expression
which gives the value of the euclidean action for theories with an
effective potentials of a rather general type,
$V(\phi,T) = a \phi^2 - b\phi^3 + c\phi^4$.  We hope that this result
will be useful for a future investigation of
bubble formation in a wide class of gauge theories with
spontaneous symmetry breaking.

On the other hand, validity of the standard assumption that the
phase transition occurs due to formation of critical bubbles should
be verified as well. Kolb and Gleiser  \cite{[7]} and, more recently,
Tetradis  \cite{Tetr} have argued that the phase transition may occur
by a different mechanism, the formation of small (subcritical)
bubbles.
If this is the case, the transition is completed earlier and by a
different mechanism than in the conventional picture.  While this
idea
is very interesting, we will argue that it is only relevant in cases
where the transition is very weakly first order and the euclidean
action corresponding to critical bubbles is not much larger than one.
This is not the case for the strongly first order phase transitions,
where the relevant value of the euclidean action at the moment of
the transition is $S \sim 130 - 140$.

Determination of the baryon asymmetry produced at the phase
transition
requires knowledge not only of how bubbles are produced, but
also of how they evolve.  Because the expansion of the universe
is so slow at this time, a typical bubble grows to a macroscopic
size before colliding with other bubbles.
In the first scenarios proposed for the formation of the asymmetry,
baryon number was produced in the bubble wall  \cite{5} --
\cite{11}.
This mechanism, at best, is not terribly efficient, because the
baryon number violating processes turn off rapidly as the scalar
field
expectation value turns on. The resulting asymmetry is sensitive to
the speed and thickness of the bubble.\footnote{The extent of this
suppression and the precise form of this dependence is still
a subject of debate.} The most effective scenarios for electroweak
baryogenesis have the baryons produced in front of the wall, in the
symmetric phase  \cite{12}.
In this picture, scattering, for example, of top quarks
from the bubble wall leads to an asymmetry in left vs. right-handed
top quarks in a region near the wall.  This asymmetry, resulting from
an asymmetry between reflection and transmission of different quark
helicities at the wall, biases the rate of baryon number violation
in the region in front of the wall; the resulting value of
$n_b /n_{\gamma}$ can be as large as $10^{-5}$. However, the authors
of Ref. \cite{12} assumed that the wall was rather thin, with a
thickness of order $T^{-1}$. For thicker walls, the
asymmetry goes rapidly to zero.  This can easily be understood.
In order to have an asymmetry in reflection coefficients,
the top quarks must have enough energy to pass through the wall.
For $m_t \sim 120$ GeV, this means typically the energy must
be greater than about $T/2$.  If the wall is very thick compared
to this scale, the motion of the top quarks is to a good
approximation semiclassical, and the reflection coefficient is
exponentially suppressed. The analyses of other authors also exhibit
sensitivity to the wall shape and velocity.

Clearly, then, it is important to understand how the bubble
propagates after its initial formation.  In this paper  we will
consider some aspects of this problem.  A complete
description of the wall evolution is rather complicated.  We will
see, however, that in certain limits it is not too difficult
to determine how the velocity and thickness of the wall depend
on the underlying model parameters.  In this analysis, it is crucial
to recall that in addition to the various microscopic parameters,
there is another parameter of great importance:  the expansion
rate of the universe. Imagine a world with arbitrarily small value of
Newton's constant. In such a world, the phase transition occurs at a
temperature arbitrarily close to $T_c$ and the pressure difference
on the two sides of the wall is arbitrarily small.  Thus, in this
limit one expects that the wall will move arbitrarily slowly, and
a systematic expansion of the relevant physical quantities in powers
of the wall velocity should be possible.  Since in the
real world, the expansion rate of the universe is indeed quite
small at $T_c$, it seems plausible that the real velocity of the
wall should be small, and that the expansion in powers of velocity
should be a fair approximation.  Even within the framework of this
small velocity approximation, we will content ourselves with
a quite crude picture for the processes which damp the wall's
motion.  Within this framework,
we will find that for plausible values of Higgs masses,
the wall velocity is indeed small, varying from $v \sim .01$
to $v \sim 0.3$; at the same time the wall will typically
have a thickness of order $10$'s of $T^{-1}$'s.

There have been a number of attempts in the past to compute the wall
parameters, including efforts by some of the present authors.
In Ref.  \cite{[3]} a simple formula for the wall velocity was given,
based on a semiclassical picture in which one species of particles
gains a large mass $M \gg T$ as it passes through the wall.
Balancing the force on the wall due to these particles with the
pressure difference between the two phases gives a relation of the
form
\beq
v = {\Delta p \over \Delta\rho} \ ,
\label{1}
\eeq
where $p$ is the pressure and $\rho$ is the internal energy.
We will see that an expression of this form holds only if the mean
free paths of the particles are large compared to the thickness of
the wall, and that this condition is not likely to hold for the
electroweak transition.  We will argue that the actual wall velocity
is somewhat larger than would be expected from this formula.

In Ref.  \cite{9}, a different approach was adopted. Assuming the
temperature and velocity are constant across the wall, momentum and
energy conservation give an expression for the velocity of the form
\beq\label{2}
v^2 \sim {\Delta p \over \Delta \rho} \ .
\eeq
As we will see in the present investigation, however, one
cannot neglect the change in temperature and velocity of the
gas as it passes the wall. For relativistic gases, one always obtains
expressions linear in the velocity.

More recently, Turok has
argued  \cite{Turok} that these types of analyses are
incorrect.  He suggests that reflection of particles from the wall
does not slow the wall at all.  Indeed, he argues that the only
way in which the wall can dissipate energy is through rather
complicated particle production processes, suppressed by several
powers of coupling constant.  As a result, even if the phase
transition is very weakly first order, the bubble wall becomes
ultrarelativistic.  To buttress his case, Turok shows that
if the gas is everywhere in thermal equilibrium (corresponding to
the local value of the scalar field) then the force on the wall
is independent of the velocity.  This is indeed correct.  However,
all of the effects we find arise because of small, velocity-dependent
departures from equilibrium.\footnote{Turok and McLerran
have recently informed us that they have reconsidered the argument
of  \cite{Turok} and have argued that there are additional sources of
damping, including scattering of particles from the wall.}

To a large extent, our analysis  will be an extension
of the semiclassical reasoning of Ref.  \cite{[3]}.
We begin by pointing out that there is an important assumption made
more or less explicitly in all of these analyses: that
by the time the wall has grown to a macroscopic size, it
has achieved a steady state.  If this is the case, the problem
can be analyzed in the rest frame of the wall, where
the scalar field and the particle distributions are
independent of time.  We describe a simple situation (due to
L. Susskind)
where this is not the case, and which we refer to as the `snowplow.'
In this situation, there is a steady pileup of particles near the
wall. While we do not expect that precisely this snowplow phenomenon
occurs in the cases of interest, it makes clear that there are
additional, potentially important effects which must be taken into
account and which are left out of existing treatments.
We then consider there three limiting situations.
In the first of these, a typical particle does not scatter
off of other particles of the gas as it passes through the wall.
This `thin wall' case is that for which (a modification of) eq.
(\ref{1}) is valid,
and tends to give a very small value for the wall velocity.
In the second case, the wall is thick compared both to typical
mean free paths for elastic scattering and for scatterings
which change the numbers of different particle types.  In this
case, the gas is nearly in equilibrium everywhere.  The velocity
is larger than in the thin wall case by
a factor of order $\sqrt{\delta \over \ell}$, where $\ell$ is
some typical mean free path and $\delta$ is the wall thickness.
In the third situation, the wall is thick compared to
typical mean free paths for elastic scattering,
but not for scatterings which change particle number.  Thus, one has
some approximate kinetic equilibrium locally, but particle numbers
are not equilibrated in the wall.  This is the situation which
appears to have greatest relevance to the electroweak transition.
Here, phenomena similar to the snowplow effect occur, and there is
an enhancement of the density of tops, W's and Z's in the wall.
This tends to reduce the wall velocity, giving a result intermediate
between the thin and thick wall cases.

The plan of the paper is the following. In Section 2 we will describe
the phase
transition in the electroweak theory in one loop approximation. In
Section 3 we
will consider  the theory of bubble formation during the phase
transition.
Section 4 will contain a discussion of role of the higher order
corrections. In
Section 5 we will deal with the issue of subcritical bubbles.
Finally, in
Section 6, the bubble wall propagation will be considered. Details of
relevant
calculations will be contained in the Appendix.

\section{The Phase Transition}

Let us consider the form of the effective
potential at finite temperature. Contributions of particles of
a mass $m$ to $V(\phi,T)$ are proportional to $m^2\,T^2$, $m^3
\,T$ and
$m^4 \ln (m/T)$. We will assume that the Higgs boson mass
is smaller than the masses of W and Z bosons and the
top quark, $m_H < m_W, m_Z, m_t$. Therefore we will
neglect the Higgs boson contribution to $V(\phi,T)$.

The zero temperature potential, taking into account
one-loop corrections, is given by  \cite{[2]}
\beq\label{3}
V_0 = - {\mu^2\over 2}\phi^2 + {\lambda\over 4} \phi^4 +
2Bv_o^2\phi^2 - {3\over 2} B\phi^4 + B \phi^4 \ln({\phi^2\over
v_o^2}) \ .
\eeq
Here
\begin{equation}\label{4}
B = {3\over 64 \pi^2 v_o^4} (2 m_W^4 +
m_Z^4 - 4 m_t^4) \ ,
\end{equation}
$v_o = 246$ GeV is the value of the scalar field at the minimum
of $V_0$, $\lambda = \mu^2/v_o^2$,
$m^2_H = 2\mu^2$. Note that these relations between $\lambda, \mu,
v_o$ and the
Higgs boson mass $m_H$, which are true at the classical level,  are
satisfied even
with an account taken of the one-loop corrections. This is an
advantage of the normalization conditions used in  \cite{[2]}.
An expression used in  \cite{[5]} is equivalent to this expression up
to an obvious change of variables.

At a finite temperature, one should add to this expression the term
\begin{equation}\label{5}
V_T = {T^4\over 2 \pi^2} \Bigl(6I_{-}(y_W) + 3I_{-}(y_Z) -
6I_{+}(y_t)\Bigr) \ ,
\end{equation}
where $y_i = M_i\phi/v_o T$, and
\begin{equation}\label{6}
I_{\mp}(y) = \pm \int_{0}^{\infty} dx
\ x^2 \ln (1  \mp e^{- \sqrt{x^2+y^2}}) \ .
\end{equation}
The results of our work are based on numerical calculation of these
integrals, without making any specific approximations  \cite{[4]}.
However, in the large temperature limit it is sufficient to use an
approximate expression for $V(\phi,T)$  \cite{[1],[5]},
\begin{equation}\label{7}
V(\phi,T) = D (T^2 - T_o^2) \phi^2 - E T \phi^3 +
{\lambda_T\over 4} \phi^4 \ .
\end{equation}
Here
\begin{equation}\label{8}
D = {1\over 8v_o^2} ( 2 m_W^2 +
m_Z^2 + 2 m_t^2) \ ,
\end{equation}
\begin{equation}\label{9}
E =  {1\over 4\pi v_o^3} ( 2 m_W^3 +
m_Z^3) \sim 10^{-2} \ ,
\end{equation}
\begin{equation}\label{10}
T^2_o = {1\over 2D}(\mu^2 - 4Bv_o^2) =
{1\over 4D}(m_H^2 - 8Bv_o^2) \ ,
\end{equation}
\begin{equation}\label{11}
\lambda_T = \lambda - {3\over 16 \pi^2 v_o^4}
\left( 2 m_W^4 \ln{m^2_W\over a_B T^2} +
m_Z^4 \ln{m^2_Z\over a_B T^2} -
4 m_t^4 \ln{m^2_t\over a_F T^2}\right) \ ,
\end{equation}
where $\ln a_B = 2 \ln 4\pi - 2\gamma \simeq 3.91$,
$\ln  a_F = 2 \ln \pi - 2\gamma \simeq 1.14$.

To avoid misunderstandings, we should note
again that due to our choice of more convenient renormalization
conditions, the form of some of our equations is
slightly different from the form of expressions used in  \cite{[5]}.
In particular, instead of  large coefficients $c_B$ and $c_F$ in the
equation for
$\lambda_T$ in \cite{[5]}, we have smaller constants $a_B$ and $a_F$.
However, all physical results obtained by our equations coincide
with the one-loop results of Refs.  \cite{[4],[5]}.

The behavior of $V(\phi,T)$ is reviewed in Refs.  \cite{[2],sher}.
It will be useful for our future discussion to identify
several `critical points' in the evolution of $V(\phi,T)$.

At very high temperatures the only minimum of $V(\phi,T)$ is
at $\phi = 0$. A second minimum appears at $T= T_1$, where
\begin{equation}\label{12}
T^2_1 = {T^2_o \over {1 - 9 E^2/8\lambda_{T_1}D}} \ .
\end{equation}
The value of the field $\phi$ in this minimum at $T = T_1$ is equal
to
\begin{equation}\label{13}
\phi_1 = {3ET_1 \over 2\lambda_{T_1}} \ .
\end{equation}
The values of  $V(\phi,T)$ in the two minima become equal
to each other at the temperature $T_c$, where
\begin{equation}\label{14}
T^2_c = {T^2_o \over {1 -  E^2/\lambda_{T_c}D}} \ .
\end{equation}
At that moment the field $\phi$ in the second minimum
becomes equal to
\begin{equation}\label{15}
\phi_c =  {2ET_c \over \lambda_{T_c}}\ .
\end{equation}
The minimum of $V(\phi,T)$ at $\phi = 0$ disappears at the
temperature $T_o$, when the field $\phi$ in the second
minimum becomes equal to
\begin{equation}\label{16}
\phi_o = {3ET_o \over \lambda_{T_o}}\ .
\end{equation}
The results of a numerical investigation of $V(\phi,T)$
for a particular case, $m_H = 50$ GeV and $m_t = 120$ GeV,
are shown in Fig. 1.

\section{Bubble Formation}

In the previous section we noted that the two minima of
$V(\phi,T)$ become of the same depth at the temperature
$T_c$, eq. (\ref{14}). However, tunneling with formation of bubbles
of the field $\phi$ corresponding to the second minimum starts
somewhat later, and it goes sufficiently fast to fill the whole
universe with the bubbles of the new phase only at some
lower temperature T when the corresponding euclidean action
suppressing the tunneling becomes less than 130 -- 140
\cite{7,[4],[5]}. Some small uncertainty in this
number is related to the speed with which bubble walls move
after being formed (see the next section) and to the exact value
of the critical temperature, which is very sensitive to the top quark
mass and to the value of the cubic term. In this paper
(see also  \cite{[4]}) we performed a numerical
study of the probability of tunneling. Before reporting our results,
we will remind the reader of some basic concepts of the theory of
tunneling at a finite temperature.

In the euclidean approach to tunneling (at zero temperature)
\cite{[6]}, the probability of bubble formation in quantum field
theory is proportional to $\exp (-S_4)$, where $S_4$ is the
four-dimensional Euclidean action corresponding to the
tunneling trajectory. In other words, $S_4$ is the instanton action,
where the instanton is the solution of the euclidean field equations
describing tunneling. A generalization of this method for tunneling
at a very high temperature  \cite{[3]} gives the probability of
tunneling per unit time per unit volume
\begin{equation}\label{19}
P \sim A(T) \cdot \exp(- {S_3\over T}) \ .
\end{equation}
Here $A(T)$ is some subexponential factor roughly
of order $T^4$;  $S_3$ is a three-dimensional instanton action.
It has the same meaning (and value) as the fluctuation of the free
energy $F = V(\phi(\vec x),T)$ which is necessary for bubble
formation. To find $S_3$, one should first find an $O(3)$-symmetric
 solution, $\phi(r)$, of the  equation
\begin{equation}\label{20}
{d^2\phi\over dr^2} + {2\over r}\,{d\phi\over dr} = V^\prime(\phi) \
,
\end{equation}
with the boundary conditions $\phi(r=\infty) = 0$ and
$d\phi/dr|_{r=0} = 0$. Here $r = \sqrt {x^2_i}$; the $x_i$ are the
euclidean coordinates, i = 1,2,3. Then one should calculate the
corresponding action
\begin{equation}\label{21}
S_3 = 4\pi \int_{0}^{\infty} r^2 \ dr\bigl[{1\over 2}
\left({d\phi\over dr}\right)^2 + V(\phi(r),T)\bigr] \ .
\end{equation}

Usually it is impossible to find an exact solution of eq. (\ref{20})
and to calculate $S_3$ without the help of a computer. A few
exceptions to this rule are given in Refs.  \cite{[2],[3]}.
One of these exceptional cases is
realized if the effective potential has two almost degenerate minima,
such that the difference $\varepsilon$ between the values of
$V(\phi,T)$ at these minima is much smaller than the energy barrier
between them. In such a case the thickness of the bubble wall at the
moment of its formation is much smaller than the radius of the
bubble, and the action $S_3$ can be calculated exactly as a
function of the bubble radius $r$, the energy difference
$\Delta V$ and the bubble wall surface energy $S_1$:
\begin{equation}\label{22}
S_3 = - {4\pi\over 3} r^3 \Delta V + 4\pi r^2 S_1 \ ,
\end{equation}
where
\begin{equation}\label{23}
S_1 = \int_0^\infty d\phi \sqrt{2V(\phi,T)} \ .
\end{equation}
The radius of the critical bubble $r$ can be found by finding an
extremum of $S_3(r)$.  However, one must be very careful when
using these results.
Indeed, as can be easily checked, this extremum is not
a {\it minimum} of the action, it is a {\it maximum}.
(This just corresponds to the fact that critical bubbles
are unstable and either expand or contract). Similarly,
the action corresponding to the true solution of eq. (\ref{20}) will
be higher than the action of any approximate solution. As a result,
one can strongly overestimate the tunneling probability by
calculating
it outside the limit of validity of the thin wall approximation.
For example, in Ref.  \cite{[5]}  the phase transition  in the
electroweak theory with $m_H \sim 50$ GeV was studied and it was
found that it happens very soon after the temperature approaches
$T_c$, occurring due to formation of bubbles with thin walls.
According to  \cite{[5]}, this happens at $\epsilon = 1/6$,  where
$\epsilon = {{T_c^2 - T^2}\over {T_c^2 - T_o^2}}$.
However, the authors did not check the validity
of the thin wall approximation in this case. Whereas our one-loop
results for the effective potential $V(\phi,T)$ are in complete
agreement with the results of Ref. \cite{[5]}, our conclusion
concerning the bubble formation is somewhat different. As we have
already mentioned, the phase transition
in the electroweak theory is completed when the ratio $S_3/T$
becomes about 130 -- 140. Our calculations show that
this happens at $\eps\sim 1/4$.
In Fig. 1 we plot the shape of the effective potential and in
Fig. 2 the shape of the solution of eq. (\ref{20})
corresponding to $S_3/T \sim 140$ for $m_H = 50$ GeV and
$m_t = 120$ GeV.  (The results for tunneling and for the ratio
$\phi/T$ prove to be not very sensitive to the mass of the top
quark in the interval  \ $100$ GeV $ \, \,  \lapp \,  \,  m_t  \, \,
\lapp \,  \,  150$ GeV.) The effective potential $V(\phi,T)$ at
$\epsilon = 1/6$ looks very similar, but the value of  $S_3/T$ in
this case is about 300, which is 3 times larger than the result
obtained in  \cite{[5]} in the thin wall approximation.
It is clear from these figures and the results of numerical
calculations of $S_3/T$ that the thin
wall approximation is far from being applicable for the investigation
of the phase transition in the electroweak theory, unless the phase
transition is weakly first order. However, the last case is not
particularly interesting from the point of view of baryon asymmetry
generation.

We must note, that the numerical results obtained
above are modified when one takes into account higher order
corrections to the effective potential. As we will show in the
next section, these corrections change the numerical
value of the coefficient  $E$ in the cubic term in (\ref{7}).
The final numerical results of our study of the probability of
tunneling will be contained, therefore, in the next section.
Here we just wanted to show the difference between the results
of the numerical investigation of tunneling and the results
obtained in the thin wall approximation. This difference remains
large after the modification of the coefficient $E$.

We would now like to obtain an analytic estimate of the probability
of tunneling in the electroweak theory,
which can be  used for any
particular numerical values of constants $D$, $E$ and
$\lambda_T$. As shown in Ref.  \cite{[5]},
eq. (\ref{7}) in most interesting cases approximates $V(\phi,T)$
with an accuracy of a few percent. This by itself does not help very
much if one must study tunneling  anew for each new set of the
constants. However, it proves possible to reduce this
study to the calculation of one function $f(\alpha)$, where $\alpha$
is some ratio of constants $D$, $E$ and $\lambda_T$. In what
follows we will calculate this
function for a wide range of values of $\alpha$. This will make it
possible to investigate tunneling in the electroweak theory without
any further use of computers.

First of all, let us represent the effective Lagrangian $L(\phi,T)$
near the point of the phase transition in the following form:
\begin{equation}\label{24}
L(\phi,T) =  {1\over 2} (\partial_{\mu}\phi)^2  -
{M^2(T)\over 2} \phi^2 + ET \phi^3 - {\la_o\over 4} \phi^4 \ .
\end{equation}
Here $M^2(T)  = 2 D (T^2 - T_o^2)$ is the effective mass squared of
the field $\phi$ near the point $\phi = 0$, $\la_t$ is the value
of the effective coupling constant $\lambda_T$ near the point of the
phase transition  (i.e. at $T \sim T_t$, where $T_t$ is the
temperature at the moment of tunneling). With a very good accuracy,
the constants
  $\la_t,  \la_{T_1}, \la_{T_c}, \la_{T_o}$ are equal to each other.

Defining $\phi = {M^2\over 2 ET} \Phi$,
$x =  X/M$, the effective Lagrangian can be written as:
\begin{equation}\label{25} L(\Phi,T) = {M^6\over
4E^2T^2}\Bigl[{1\over 2} (\partial_{\mu} \Phi)^2  -
{1\over 2} \Phi^2 + {1\over 2} \Phi^3 - {\alpha\over 8} \Phi^4 \Bigr]
\ ,
\end{equation}
where
\begin{equation}\label{26}
\al = {\la_o M^2\over 2 E^2T^2}\ .
\end{equation}
The overall factor ${M^6\over 4E^2T^2}$ does not affect the Lagrange
equation
\begin{equation}\label{27}
{d^2\Phi\over dR^2} + {2\over R}\,{d\Phi\over dR} = \Phi -  {3\over
2}\Phi^2 +  {1\over 2}\alpha \Phi^3\ .
\end{equation}
Solving this equation and integrating over $d^3X = M^{-3} d^3x$
gives the following expression for the corresponding action:
\begin{equation}\label{28}
{S_3\over T} = {{4.85\,{M^3}}\over {{{ E}^2}\,{T^3}}} \times
f(\alpha) \ .
\end{equation}
The function $f(\alpha)$ is shown in Fig. 3. It is equal  \cite{[3]}
to 1 at  $\alpha= 0$, and blows up when $\alpha$ approaches 1.
In the interval from 0 to 0.95 this function, with an accuracy
about 1\%, is given by the following simple expression:
\begin{equation}\label{29}
f(\alpha) = 1 + {\alpha\over 4} \Bigl[ 1
+{{2.4}\over {1 - \alpha}} +
 {{0.26}\over{{{( 1 - \alpha ) }^2}}}\Bigr] \ .
 \end{equation}

In the vicinity of the critical temperature $T_o$, i.e. at  $\Delta T
\equiv T - T_o \ll T_o$, the action (\ref{28}) can be written in the
following form:
\begin{equation}\label{30}
 {S_3\over T} =  {38.8\,D^{3/2}\over {{{ E}^2}}}\cdot \left({\Del
T\over T}\right)^{3/2} \times f\Bigl({2\,\la_o\,D\,\Del T\over
E^2\,T}\Bigr) \ .
\end{equation}
Using these results, one can easily get analytical expressions for
the tunneling probability in a wide class of theories with
spontaneous symmetry breaking, including GUTs and the minimal
electroweak theory.

\section{Infrared Problems and Reliability of the Perturbation
Expansion}

In our previous discussion, we have considered only the one loop
corrections to the effective potential.  In this section we
discuss the role of higher order corrections.

Early investigations of the
electroweak phase transition \cite{[1]} did not take into account
corrections due to strong interactions, since at that time
most physicists did not expect that top quarks would be heavier
than W and Z bosons, and thus their contributions were not
expected to be terribly important.  Given what we now know about the
top quark mass, it appears that top
quarks give the largest contribution to the parameters D
and $\lambda_T$ in eqs. (\ref{8}), (\ref{11}).  Thus, it is
natural to ask whether strong interaction corrections are likely
to be important.

Our preliminary investigation of this question indicates
that this is not the case.
For example, one of the most important effects would be a
change of the Fermi distribution at a finite temperature due
to the modification of the quark mass by interactions with gluons.
According to \cite{[9]}, quarks at a high temperature
acquire a correction to their effective mass squared,
\begin{equation}\label{17}
\Delta m^2_t(T) = {g_s^2\over 6} T^2
\end{equation}
with $g_s$ the strong coupling constant.\footnote{We use
$\alpha_s \sim  0.1$ for our estimates.}
This gives $\Delta m^2_t(T) \sim  0.2 \, T^2$ at the temperature
of the electroweak phase transition, $T \sim 10^2$ GeV.
A similar contribution to the boson mass could lead to important
effects (see below). However, due to Fermi statistics, fermion
propagators contain terms $[(2n+1)\,\pi\,T]^2$ \cite{[2]}.
As a result, all thermodynamic quantities rather weakly depend
on the effective mass squared of the fermions, until this mass
becomes comparable with $\pi T$.

One should note, of course, that $\Delta m^2_t(T) \sim  0.2 \, T^2$
is not a
true correction to the top quark mass squared; rather it is a
square of the mass gap in the spectrum of fermionic excitations.
Therefore a more detailed analysis of the higher-loop diagrams
involving strong interactions is desirable.
Nevertheless, our estimate suggests that the strong interaction
effects are actually insignificant for the description of the phase
transition. They just lead to a small modification of the critical
temperature.
More importantly, they do not change cubic terms $\sim g^3 \phi^3 T$
in the effective potential and do not induce linear terms $\sim g^3
\phi T^3$.

However, higher order corrections in weak interactions at $T \geq
T_c$ may be very important.
It is well known that, in field theories of massless particles,
perturbation theory at finite temperature
is subject to severe infrared divergence problems.  For small
values of the scalar field, the gauge bosons (and near the
phase transition, the Higgs boson) are nearly massless;
as a result, as was pointed out in the early work on this subject
\cite{[1b],[11]}, one cannot reliably compute the effective
potential for very small $\phi$.  In this section, we attempt to
determine whether the standard one loop calculation of $V$ is indeed
reliable in the range of $\phi$ relevant to our analysis.  One might
worry, for example, that since the term $- E T \phi^3 $
in the effective potential, which leads to the  first order phase
transition, is important only for rather small $\phi$, there might
be large corrections
changing the order of the phase transition.  We will show
that indeed the coefficient of the $\phi^3$ term is altered
to $2/3$ of its one loop value.  This renders the phase transition,
for a given value of the couplings, less first order and
can have significant effects on baryogenesis.
On the other hand, we will
argue that perturbation theory is not in terribly bad shape,
and that one can determine the nature of the phase transition
with some confidence from low order calculations.

Recently, in a very interesting paper, Brahm and Hsu reach the
opposite conclusion \cite{Hsu}. These authors find that
at small $\phi$, higher order corrections to the scalar
field contribution to the effective potential contain a
large negative linear term $-\, g^3\phi T^3$, which eliminates
any trace of a first order transition.
They argue, moreover, that their calculation is reliable,
{\em i.e.} that all other higher order corrections are under
control and do not modify their conclusion.

On the other hand, Shaposhnikov considers higher order
corrections to the vector particle contribution to $V(\phi,T)$
and finds a large positive term $+\, g^3\phi T^3$. He
concludes that the phase transition is strongly first order
($\phi/T\, >\, 1$) even for $m_H \sim 64$ GeV \cite{Shap}.

We will show that neither positive nor negative
linear terms appear  in the expression for
$V(\phi,T)$ if one studies higher order corrections paying
particular attention to the correct counting of Feynman diagrams.
\cite{OurPaper}. We will employ two separate approaches to this
problem: a straightforward enumeration of Feynman diagrams,
and an effective
action analysis valid for a discussion of infrared effects.

We will consider here for simplicity the contribution of the scalar
particles and the W bosons
only; adding the contribution of Z bosons is trivial.
As we have already noted, for questions of infrared behavior,
fermions may be ignored. Coulomb gauge,
$\vnab\cdot\vec W = 0$, is particularly convenient for the analysis,
though the problem can be analyzed in other gauges as well. In this
gauge, the vector field propagator
$D_{\mn}$ after symmetry breaking (and after a proper
diagonalization) splits into two pieces, the
Coulomb piece, $D_{00}$, and the transverse piece, $D_{ij}$.  For
non-zero values of the discrete frequency, $\omega_n=2\pi n T$,
the Coulomb piece mixes with the 'Goldstone' boson. However,
for the infrared problems which concern us here, we are only
interested in the propagators at zero frequency.  For
these there is no mixing.  One has \cite{[1b]}
\begin{equation}\label{i1}
D_{00}(\omega=0,\vk) = {1 \over \vk^2 + m^2_W(\phi)}
\end{equation}
and
\begin{equation}\label{i2}
D_{ij}(\omega=0,\vk) = {1 \over\vk^2 + m^2_W(\phi)}P_{ij}(\vk) \ ,
\end{equation}
where $P_{ij}=\delta_{ij} -{k_i k_j \over \vk^2}$. The mass of the
vector field $W$ at the classical level is given by $m_W = g v_o /2$.
Propagators of the Higgs field $\phi$ and of the 'Goldstone' field
$\chi$ in this gauge are given by
\begin{equation}\label{i3}
D_\phi(\vk) = {1 \over  \vk^2 + m^2_\phi} \ ,
\end{equation}
\begin{equation}\label{i4}
D_\chi(\vk) = {1 \over \vk^2} \ .
\end{equation}

Let us review several ways of obtaining the standard one-loop
expression for the cubic term in the effective potential, eq.
(\ref{7}). The most straightforward
is to carefully expand eq. (\ref{5}) for the effective potential in
$y_W = {m_W\over v_o T} = {g \phi\over 2T}$. Indeed, the contribution
of $W$-bosons to the effective potential at $T > m_W(\phi)$ is given
by
\begin{eqnarray}\label{i7}
V_W(\phi,T) &=& 2\times 3\times \Bigl( - {\pi^2\over 90} T^4 + {
m^2_W(\phi)\over 24} T^2 - {m^3_W(\phi) \over 12 \pi} T
+ \cdots\Bigr) \nonumber\\
  &=& 2\times 3\times \Bigl( - {\pi^2\over 90} T^4 + {
g^2 \phi^2\over 96} T^2 - {g^3\phi^3 \over 96\pi} T
+ \cdots\Bigr)
\ .
\end{eqnarray}
Here the expression in brackets coincides with the contribution of a
scalar field
with mass $m_W$; the factor $2$ appears since there are two
$W$-bosons with opposite charges, while
the factor 3, which will be particularly important
in what follows, corresponds to the two transverse and one
longitudinal degrees of freedom with mass $m_W$.

Alternatively, we can obtain the cubic term by looking
directly at the one-loop Feynman diagrams.  For this purpose,
it is only necessary to examine the zero frequency contributions.
Certain diagrams containing {\it four} external lines of the
classical scalar field naively give a contribution proportional to
$g^4 \phi^4$; the cubic term arises because the zero
frequency integrals diverge for small mass as $T / m_W
\sim T / g \phi$.

Consider, in particular, the zero frequency part of the expression
for the one loop free energy in momentum space.  It is simplest
to compute the tadpole diagrams for $dV / d\phi$, as indicated
in Fig. 4, and afterwards integrate with respect to $\phi$.
The transverse gauge bosons give a contribution
\begin{equation}\label{trzero}
{dV_{{\rm tr}}\over d\phi} =
 2\times  {g^2 \phi T \over 2}\ \int {d^3 k \over (2 \pi)^3}
 {1\over \vk^2 + m_W^2}
   = -2\times {g^2\phi\; T\over 8\pi}\, \sqrt{m_W^2} \ ,
\end{equation}
where, by keeping only the zero frequency mode, we have
dropped terms which are analytic in $m^2$.\footnote{ The integral
has been defined by dimensional regularization; ultraviolet
divergences here are absorbed into the usual zero temperature
renormalizations.} The Coulomb lines give half the result of
eq. (\ref{trzero}). Integration of the total vector field
contribution correctly represents the cubic term in (\ref{i7}).

A complete gauge boson contribution to the tadpole, including
  the non-zero frequency modes, is \cite{[1b]}
\beq\label{i9}
{dV_W(\phi,T)\over d\phi}  = 2\times 3\times  {g^2\phi\over 48} \
\Bigl( T^2 -
{3 m_W T \over\pi }+ \cdots\Bigr) = 2\times 3\times  {g^2\phi\over
48} \
\Bigl( T^2 - {3\,g\phi T \over 2 \pi }+ \cdots\Bigr) \ .
\eeq
One can easily check that integration of this expression with respect
to $\phi$ gives eq. (\ref{i7}).

With these techniques, we are in a good position to study higher
order corrections to the potential.
The authors of Refs. \cite{Hsu}, \cite{Shap} found a linear
contribution to the potential by substituting the mass found at one
loop back into the one loop calculation.  The effective
masses-squared of both scalar particles and of the Coulomb field
contain terms of the form $\sim g^3 T \phi$, which, upon substitution
in (34), give linear terms. But this procedure  is not always
correct.
It is well known that the sum of the geometric progression,
which appears after the insertion of an arbitrary number of
polarization operators $\Pi(\phi,T)$ into the propagator
$ (k^2 + m^2)^{-1}$, simply gives $(k^2 + m^2 + \Pi(\phi,T))^{-1}$.
Therefore one can actually use propagators
$(k^2 + m^2 + \Pi(T))^{-1}$, which contain the effective mass-squared
$m^2 + \Pi(\phi,T)$ instead of $m^2$. However, this trick
with the geometric progression does
not work for the closed loop diagram for the effective potential,
which contains $\ln (k^2 + m^2)$. A naive substitution of the
effective mass squared $m^2 + \Pi(\phi,T)$ instead of $m^2$ into
$\ln (k^2 + m^2)$ corresponds to a wrong counting of
higher order corrections.

A simple way to take into account high temperature corrections
to masses of vector and scalar particles without any
problems with combinatorics
is to compute tadpole diagrams for ${d\,V\over d\phi}$; these
are then trivially integrated to give the potential.
One can easily
check by this method  that  no linear terms appear
in the expression for  $V(\phi,T)$.
Indeed, at a given temperature and
effective mass, the tadpoles are linear in $\phi$ (see e.g. equation
(\ref{i9})). To take into
account the mass renormalization in the tadpoles, one
 should substitute the effective mass squared
$m^2 + \Pi(\phi,T)$ into the one-loop expression for the tadpole
contribution; as we explained above (see also (\cite{[1b]})),
this is a correct and unambiguous procedure for tadpoles.
Since $m^2 + \Pi(\phi,T)$ is not singular in the limit
$\phi\rightarrow 0$, the tadpole (\ref{i9}) in this limit remains
linear in $\phi$. Therefore its integration with respect
to $\phi$, which gives the
correction to the effective potential,  is quadratic  in $\phi$,
{\it i.e.} it does not contain any linear terms.
\footnote{The absence of a linear term in $\phi$ can be
understood in an effective
Lagrangian approach as well.  Such an approach automatically
gives the correct combinatorics. Since, at weak coupling, we
are interested in energy scales much less than $\pi T$,
one can first integrate out all modes of the various fields with
non-zero Matsubara frequency, thereby obtaining
a three-dimensional effective action for the light fields.
At one loop, one has the usual mass corrections,
order $g^2T^2$ for the Coulomb and scalar lines, whereas the
quadratic term for the transverse gauge bosons goes to zero with
$\vk$ and $\phi$. These `mass corrections' are analytic in
$|\phi|^2$;
in particular, they do {\em not} contain linear terms in $\phi$.
By assumption, the Higgs field is light at the
phase transition, but one can again integrate out the massive
Coulomb field.  It is still true (as a consequence of
gauge invariance) that the quadratic term for the transverse
gauge bosons
vanishes at zero momentum. Having done this procedure, we compute the
effective potential for $\phi$ by calculating the determinant of
Gaussian fluctuations in the low energy theory, i.e. the three
dimensional theory containing only transverse gauge bosons and
scalars. The linear terms in $\phi$ used in Ref. \cite{Shap}
come from loops of {\em light} fields ($\om=0$), and so in the
present language we need to check that there is no non-analytic
behavior in higher loop graphs in the effective theory.
In fact, although individual diagrams may be singular, the
sum of all two loop graphs is non-singular, essentially due to
gauge invariance, and so the effective potential contains no
linear term in $\phi$, to this level of approximation. The cubic
term is correctly reproduced by this process (see below).}

Even though there are no linear terms $\sim g^3 \phi T^3$, higher
order corrections do have a dramatic effect
on the phase transition, which has apparently not been noted
before. This effect is a modification of the cubic term.

As we have shown above, the cubic term appears due to the
contribution of zero modes, $\om_n = 2\pi nT = 0$. This makes it
particularly easy to study its modification by high order effects.
Indeed, it is well known that the Coulomb field at zero frequency
acquires the Debye `mass',  $m^2_D = \Pi_{00}(\om_n = 0, \vk \to 0)
\sim g^2 \, T^2$.
This leads to an important modification of the Coulomb
propagator (\ref{i1}):
{\begin{equation}\label{coulombloop}
D_{00}(\vk\to 0)={1 \over \vk^2 + m_D^2 + m_W^2(\phi)} \ .
\end{equation}
For the values of $\phi$ of interest to us, $m_D^2 \gg m_W^2(\phi)$.
Thus, repeating the calculation of the cubic term,
the Coulomb contribution disappears.  However, the transverse
contribution, which is two times larger than the Coulomb one,
is unaffected at this order, due to the vanishing of the `magnetic
mass'  \cite{[11],LGPY}.  As a
result, the  cubic term does not disappear,  but it is diminished by
a factor\footnote{We should point out that if the magnetic mass of
the transverse gauge bosons were to be numerically rather large,
then the factor $E$ would be reduced further, leading to a further
weakening of the order of the phase transition.}
 of $2/3$:
 \footnote{After we obtained this result, we received a
paper by Carrington \cite{Carr} where the renormalization of
the cubic term was also considered. A similar investigation was
performed by Shaposhnikov as well \cite{Shap}. These authors did
not point out that they obtained the renormalization of the cubic
term by the factor of $2/3$, but after some algebra one can identify
those terms in their expressions which are equivalent to ours.
However, for different reasons, the final numerical results for the
ratio $\phi/T$ obtained by both Shaposhnikov and Carrington
differ from our result considerably, being approximately two
times larger than our result shown in Fig. 5.}
\begin{equation}\label{newE}
E =  {1\over 6\pi v_o^3} ( 2 m_W^3 +
m_Z^3) \ .
\end{equation}

 This small correction proves to be very significant. Indeed,
eqs. (\ref{15}), (\ref{16}) show that the ratio of the scalar field
$\phi$ to the
temperature at the moment of the phase transition is proportional to
$E$, i.e.
to the cubic term. Actually, the dependence is even slightly
stronger, since
for smaller  E the tunneling occurs earlier. Results of a complete
numerical
investigation of the ratio $\phi/T$ at the moment of the phase
transition, as a
function of the Higgs boson mass are shown in Fig. 5, for the top
quark mass $ m_t = 120$ GeV. We have found that the ratio
$\phi/T$ is not very sensitive  to the mass of the top quark, in the
interval $100$ GeV $\, \,  \lapp \,  \,  m_t\, \,  \lapp \,  \,  150$
GeV, and it
decreases for $m_t$ outside this interval.  Even before
the reduction of the cubic term  was taken into account, the ratio
$\phi/T$ for $m_H \, \, \gapp \, \, 57$ GeV was slightly less than
the critical
value $\phi/T \approx 1$. The
decrease of this quantity by a factor of  $2/3$ makes it absolutely
impossible to preserve the
baryon asymmetry generated during the phase transition in the minimal
model of electroweak interactions with $m_H \, \,  \gapp \, \, 57$
GeV.

Is this the end of the story? The effective coupling constant of
interactions
between W bosons and Higgs particles is $g/2$. In this case, a
 general investigation of the infrared
problem in the non-Abelian gauge theories at a finite temperature
suggests that  the results which
we obtained are reliable for
$\phi \  \,  \gapp \,  \ {g\over 2} \ T \sim T/3 $ \cite{[11]},
\cite{LGPY}.
Thus,  a more detailed investigation is needed to study behavior
of the theories  with $m_H \, \,  \gapp \, \,  10^2$ GeV near the
critical temperature, since the scalar field,
which appears at the moment of the phase transition
in these theories, is very small (see Fig. 5). However, we expect
that our results are reliable for strongly first order phase
transitions with $\phi\,\, \,  \gapp \, \,\, T$, which is quite
sufficient
to study (or to rule out) baryogenesis in the electroweak theory.

Finally, we would like to address a fundamental question:  since
the theory for $\phi\ll gT$ is infrared divergent, can we definitely
establish that the symmetry is restored at high temperature,
or is it possible that $\phi$ always has some small, non-zero
value?  To address this question, we can work at $T\gg T_o$.
In this case, the scalar field is massive, and scalar loops
are not singular in the infrared. Potential infrared problems
arise only from gauge boson loops. For these, the situation
is similar to that in high temperature QCD \cite{LGPY}.
The standard assumption about QCD is that the infrared divergences
are cut off at a scale $m_{mag}\sim g^2 T$
(the detailed mechanism of the infrared cutoff
will not be important to us).
The free energy, $\Om$, is non-singular through order $g^4\, T^4$.
At order $g^6\, T^4$, there is a logarithmic infrared divergence;
$\Om\sim g^6 \, T^4 \ln (m_{mag}/T) \sim g^6\, T^4 \ln g^2$.
Higher order corrections go as $g^6\,T^4 (g^2\, T / m_{mag})^n$,
{\em i.e.} they are all of the same
order. It is usually assumed, then, that the free energy can
be computed through order $g^6\, T^4 \ln g^2$.
A similar investigation suggests that the effective mass can be
calculated with an accuracy $g^4\, T^2 \ln g^2$.

What are the consequences of these assumptions for the electroweak
theory? First, the existence of an infrared cutoff of order $g^2\, T$
means that the potential is analytic in $|\phi |^2$,
for small $\phi$. This implies, in particular, that there are no
linear terms in $\phi$. Consider, then, the calculation of the
$\phi^2$ term, which determines the value of the critical
temperature. At lowest order, one has the standard result,
$D T^2  \phi^2 \sim g^2 T^2 \phi^2$.  At two loop
order, there is a correction proportional to
$g^4 T^2 \phi^2 \ln g^2$; higher order corrections all
go as $g^4 T^2 \phi^2$.  Thus, as in QCD, we can say that
the mass term can be calculated to order $g^4 T^2 \ln g^2$.
This is a small correction, and the one-loop calculation
is reliable: for small $\phi$, the curvature of
the potential is the sum of the (zero-temperature)
negative term $- {\mu^2\over 2}\phi^2  + 2Bv_o^2\phi^2$,
and a positive term $D T^2 \phi^2 + O(g^4 \ln g^2) T^2 \phi^2$
which grows with temperature. Thus, we have a phase transition,
and the higher loop effects can only lead to corrections
$\sim g^2 \ln g^2$ to the value of the critical  temperature
obtained in the one loop approximation.

Note that this discussion is valid for any value of the
Higgs mass. If these arguments are correct, then we expect
the situation with the phase transitions in the non-Abelian
gauge theories to be the same as in the standard
case: infrared problems may prevent a simple description of the
phase transition in a small vicinity of the
critical point (unless the phase transition is strongly first order),
but everywhere outside this region, the symmetry behavior
of gauge theories can be described in a reliable way.
We hope to return to a discussion of this interesting question
in a separate publication.

\section{Subcritical Bubbles}

Despite our semi-optimistic conclusions concerning the
infrared problem, it is still desirable to check that the whole
picture of the behavior of
the scalar field described above is (at least) self-consistent. This
means that
if the effective potential is actually given by eqs. (\ref{7}),
(\ref{8}),
(\ref{10}), (\ref{11}), (\ref{newE}), then our subsequent description
of the
phase
transition and
the bubble formation is correct. Indeed, one would expect that the
theory of
bubble formation is reliable, since the corresponding action for
tunneling
$S_3/T $ is very large, $S_3/T \sim 130 - 140$.  However, recently
even the validity of this basic assumption has been questioned.
Gleiser and Kolb \cite{[7]} and Tetradis  \cite{Tetr} have argued
that in many cases phase transitions occur not  due to bubbles of a
critical size, which we studied in section 3, but due to
smaller, subcritical bubbles.  We believe that these
authors raise a real issue.
However, we will now argue that this problem only arises if the
phase transition is extremely weakly first order.

The basic difference between the analysis of Ref.
\cite{[7],Tetr} and the more conventional one is
their assumption that at the time of the phase transition
there is a comparable probability to find different parts of the
universe in either of the two minima of $V(\phi,T)$. The main
argument of Ref. \cite{[7],Tetr} is that if the dispersion
of thermal fluctuations of the scalar
field $<\phi^2>\sim T^2$ is comparable with the distance between the
two minima of $V(\phi,T)$, then the field $\phi$ ``does not know''
which minimum is true and which is false. Therefore it spends
comparable time in each of them. According to  \cite{[7]},  a
kind of equilibrium between the
domains of the two types is achieved due to subcritical bubbles with
small action $S_3/T$ if many such bubbles may appear within a
horizon of a radius $H^{-1}$.

In order to investigate this question in a more detailed way, let us
re-examine our own assumptions concerning the
distribution of the scalar field $\phi$ prior to the moment at which
the temperature drops down to $T_1$, when the second minimum of
$V(\phi,T)$  appears. According to (\ref{13}), the value of the
scalar field
$\phi$ in the second minimum at the moment when it is formed is equal
to $\phi_1 = {3ET\over 2\lambda_T}$.  For $m_H \sim 60$ GeV
(and taking into account the coefficient $2/3$ in the cubic term) one
obtains $\phi_1 \sim 0.4\, T$. Thermal fluctuations of the
field $\phi$ have the dispersion squared $<\phi^2> = T^2/12$.
(Note an important factor $1/12$, which was absent in the
estimate made in \cite{[7]}.) This gives dispersion of thermal
fluctuations $\sqrt{<\phi^2>} \sim 0.3\, T$,
which is not much smaller than $\phi_1$.

However, as the authors of  \cite{[7]} emphasized in their
previous work  \cite{[12]} (see also  \cite{Tetr}), the total
dispersion $<\phi^2> \sim T^2/12$
is not an adequate quantity to consider since we are not really
interested in infinitesimally small domains containing different
values
of fluctuating field $\phi$. They argue that the proper measure of
thermal fluctuations is the contribution to $<\phi^2>$ from
fluctuations of the
size of the correlation length $\xi(T) \sim M^{-1}(T)$. This leads to
an
estimate $<\phi^2> \sim T\,M(T)$, which also may be quite large
 \cite{[12]}. Here again one should be very careful to use the proper
coefficients in the estimate. One needs to understand also why this
estimate could be relevant.

In order to make the arguments of Ref.  \cite{[7],Tetr} more
quantitative
and to
outline the domain of their validity, it is helpful to review the
stochastic approach to tunneling (see  \cite{[10]} and
references
therein). This approach is not as precise as the euclidean approach
(in theories where the euclidean approach is applicable). However,
it is much simpler and more intuitive, and it may help us to look
from
a different point of view on the results we obtained in the previous
section and on the approach suggested in  \cite{[7],Tetr}.

The main idea of the stochastic approach  can be illustrated by an
example of tunneling with bubble formation from the point
$\phi = 0$ in the theory  (\ref{24})  with the effective potential
\begin{equation}\label{31}
V(\phi,T) = {M^2(T)\over 2} \phi^2 - ET\phi^3  +  {\la_o\over 4}
\phi^4 .
\end{equation}
For simplicity, we will study here the limiting case $\la_o\to 0$.

At the moment of its formation, the bubble wall does not move.
In the limit of small bubble velocity, the equation of motion of the
field $\phi$ at finite temperature is simply,
\begin{equation}\label{32}
\ddot\phi = d^2\phi/dr^2 + (2/r)d\phi/dr  - V^\prime(\phi) \  .
\end{equation}
The bubble starts growing if $\ddot\phi > 0$, which requires that
\begin{equation}\label{33}
 |d^2\phi/dr^2 + (2/r)d\phi/dr| < - V^\prime(\phi) \ .
\end{equation}
A bubble of a classical field is formed only if  it
contains a sufficiently big field $\phi$. It should be over the
barrier,  so that
$dV/d\phi < 0$, and the effective potential there should be
negative since otherwise formation of a bubble will be
energetically unfavorable. The last condition means
that the field $\phi$ inside the critical bubble should be
somewhat larger than  $\phi_*$, where $V(\phi_*,T) = V(0,T)$.
In the theory (\ref{31}) with $\la_o\to 0$, one has
$\phi_* = M^2/2ET$. As a simplest (but educated) guess,
let us take  $\phi \sim  2\phi_* = M^2/ET$.
Another important condition is that the size of the bubble
should be sufficiently large. If the size of the bubble
is too small, the gradient terms are bigger than the term
$|V^\prime(\phi)|$, and the field
$\phi$ inside the bubble does not grow. Typically, the second term in
 (\ref{33})
somewhat compensates the first one. To make a very rough estimate,
 one may write the condition (\ref{33}) in the form
\begin{equation}\label{34}
{1\over 2} r^{-2} \sim
{1\over 2} k^2 <  {1\over 2} k^2_{max}  \sim \phi^{-1}
|V^\prime(\phi)| \sim 2 M^2  .
\end{equation}
Let us estimate the probability of an event in which thermal
fluctuations
with $T \gg M$ build up a configuration of the field
satisfying this condition.
The dispersion of thermal fluctuations of the field
$\phi$ with $k<k_{max}$ is given by
\beqra\label{35}
<\phi^2>_{k<k_{max}} & = &
{1\over 2\pi^2}\int_{0}^{k_{max}}{k^2 dk\over
\sqrt {k^2 + M^2}\left(\exp{\sqrt{k^2 + M^2(\phi)}\over T} - 1
\right)} \nonumber \\
 & \sim & {T \over 2\pi^2}\int_{0}^{k_{max}}{k^2 dk\over
{k^2 + M^2}} \ .
\eeqra
Note that the main contribution to the integral is given
by $k^2 \sim k^2_{max} \sim 4 M^2 $.
This means that one can get a reasonably good estimate
of $<\phi^2>_{k<k_{max}}$ by omitting  $M^2$ in the
integrand. This also means that this estimate will be good
enough even though the effective mass  of the scalar field
$M^2(\phi)=V^{\prime\prime}(\phi)$ changes
between $\phi = 0$ and
$\phi$. The result we get is
\begin{equation}\label{36}
<\phi^2>_{k<k_{max}} \simeq    {T\over 2\pi^2}
\int_{0}^{k_{max}}{ dk}
  = { T k_{max}\over 2\pi^2} = {C^2   TM\over  \pi^2} \ .
\end{equation}
Here $C = O(1)$ is a coefficient reflecting the uncertainty
in the determination of  $k_{max}$ and estimating the integral.

Thus, we have a rough estimate of the dispersion of
perturbations which may  sum up
to produce a  field $\phi$ which satisfies the condition (\ref{34}).
We can  use it to evaluate the probability
that these fluctuations build up a bubble of the field $\phi$ of a
radius $r > k^{-1}_{max}$. This can be done with the help of the
Gaussian
distribution\footnote {The probability distribution is approximately
Gaussian even though the effective potential is not purely quadratic.
The reason is that we were able to neglect the curvature of the
effective potential {$m^2 = V^{\prime\prime}$} while calculating
{$<\phi^2>_{k<k_{max}}$.}}
\begin{equation}\label{37} P(\phi)
\sim \exp(-{\phi^2\over 2<\phi^2>_{k<k_{max}}}) =   \exp (-{
M^3\pi^2\over  2 C^2 E^2 T^3}) \sim \exp (-{4.92 M^3\over C^2
E^2 T^3}) \ .
\end{equation}
Note that the factor in the exponent in (\ref{37}) to within  a
factor of $C^2 = 1.02$  coincides with the exact result for  the
tunneling probability in this theory obtained by the euclidean
approach
 \cite{[3]} (see eq. (\ref{29})):
\begin{equation}\label{38}
P \sim \exp (-{4.85 M^3\over  E^2 T^3}) \ .
\end{equation}
Taking  into account  the very
rough method we used to calculate the dispersion of the perturbations
responsible for tunneling, the coincidence is rather impressive.

As was shown in \cite{[10]}, most of the results obtained in
the tunneling theory by
euclidean methods can easily be reproduced (with an accuracy  of the
coefficient  $C^2 = O(1)$ in the exponent) by this  simple method.

Now let us return to the issue of subcritical bubbles. As we have
seen, dispersion of the long-wave perturbations of the scalar field,
$<\phi^2>_{k<k_{max}} \simeq {k_{max}\,T\over 2\,\pi^2}$,
is quite relevant  to the theory of tunneling. Its calculation
provides a
simple and intuitive way to get the same results as we obtained
earlier by the euclidean approach  \cite{[10]}.  To get a good
estimate of the
probability of formation of a critical bubble in our simple model one
should calculate  this dispersion for $k_{max} \sim 2 M(T)$, which
gives $<\phi^2>_{k<k_{max}} = TM/\pi^2$. Note, that this
estimate is much
smaller than the naive estimate $<\phi^2> \sim TM$.

The crucial test of our basic assumptions  is a comparison
of this dispersion and the value of the field $\phi$ at
the moment $T = T_1$, when the minimum at $\phi   =
\phi_1 \not = 0$ first appears. Using eqs. (\ref{7}),
(\ref{13}), one can easily check that the mass of the
scalar field at $T = T_1$, $\phi = 0$ is given by
\begin{equation}
m = {3ET\over 2 \sqrt\lambda_T} \ .
\end{equation}
This yields
\begin{equation}
 \sqrt {<\phi^2>}_{k<k_{max}} \  \sim
 \phi_1\  {\lambda^{3/4}\over\pi\sqrt {3E/2}} \approx \phi_1 \
{10  \lambda_T^{3/4}\over  \pi } \ .
\end{equation}
For the Higgs boson with $m_H  \sim 60$ GeV one obtains
\begin{equation}
\sqrt {<\phi^2>}_{k<k_{max}} \ \sim   {\phi_1\over 5} \ .
\end{equation}
Thus, even with account taken of the factor $2/3$ in
the expression for $E$, the dispersion of long-wave fluctuations
of the scalar field is much smaller than the distance
between the two minima. Therefore,
the field $\phi$ on a scale equal to its correlation length
$\sim M^{-1}$ is not equally distributed between the two
minima of the
effective potential. It just fluctuates with a very small amplitude
near the point $\phi = 0$.
The fraction of the volume of the universe filled by the field
$\phi_1$ due to these fluctuations (i.e. due to subcritical bubbles)
for $m_H  \sim 60$ GeV is negligible,
\begin{equation}\label{ttt} P(\phi_1)
\sim\exp\left(-{\phi^2\over 2<\phi^2>_{k<k_{max}}}\right)
\sim\exp\left(-{3E\,\pi^2\over 4\la_T^{3/2}}\right)  \sim e^{- 12} \
{}.
\end{equation}
Since we already successfully applied this method for
investigation of tunneling, we expect that this estimate
is also reliable. The answer  remains rather small even
for $m_H \sim 100$ GeV, when
the phase transition is very weakly first order.

Moreover, even these long-wave fluctuations do not lead to
formation of stable
domains of space filled with the field $\phi \not = 0$,
until the temperature is below $T_c$ and
critical bubbles appear.
One expects a typical subcritical bubble to
collapse in a time $\tau \sim k_{max}^{-1}$; this is about
thirteen orders of magnitude smaller than
the total duration of the phase
transition, $\Delta t \sim 10^{-2} H^{-1} \sim 10^{-4} M_p \,
T^{-2}$.
We do not see any mechanism which might increase $\tau$ by
such a large factor;
effects such as
decrease of the speed of the collapse of such
bubbles due to finite-temperature effects (considered in the next
section), or the inefficiency of radiating away  the energy of
oscillating
subcritical bubbles  \cite{[13]} are much more modest.

Despite all these comments, we think that  subcritical
bubbles deserve further investigation.  They may lead to interesting
effects during phase transitions in GUTs, since the difference
between $T^{-1}$ and the duration of the GUT phase transitions
is not as great
as in the electroweak case. They may play an important role in  the
description of the electroweak phase transition as well, in models
where the phase transition occurs during a time not much
longer than $T^{-1}$. This may prove to be the case for very weakly
first order phase transitions with $m_H \, \,  \gapp \, \, 10^2$ GeV,
when the
distance between the two minima of $V(\phi,T)$ at $T \sim T_1$
is smaller than the dispersion
$\sqrt{<\phi^2>}_{k<k_{max}}\sim\sqrt{TM}/\pi$.

\section{Propagation of the Bubble Wall}

\subsection{General Considerations}

After bubbles are formed, one expects that they will grow until
they collide.  Since the expansion rate is so small at this time,
provided the bubbles have a velocity which is not {\it extremely}
small, typical bubbles will grow to a macroscopic size.
Thus, it is important to understand how bubbles
propagate long after their formation. An underlying assumption in
discussions of the evolution of the bubble wall to date is that
some time after the formation of the bubble, a steady state
situation is achieved, in which the scalar field, temperature,
and particle velocities are all constant in time in a
frame which we will call the `wall frame.'  While it is
plausible that such a state is achieved, we will not prove
that this is the case; indeed, as we explain below, one can
imagine situations in which this is not the case.  To develop some
intuition, we will consider two simple models.
One picture, which suggests that a steady state will arise, is the
following. As the wall passes through
the medium, it dissipates energy and heats the gas.
Since the wall is quite permeable, especially to light quarks,
heat is readily transported both behind and in front of the wall.
Once the wall is
very large, the problem becomes one dimensional, so consider
as a model a point source of heat, moving in an incompressible fluid
 with velocity
$v$.  The temperature will obey a diffusion equation, with
diffusion coefficient $\chi$. The solution of this equation is
\begin{equation}\label{40}
T-T_{\infty}= {q \over v \sqrt{\chi}} \int^0_{-\infty}{dx^{\prime}
\over
\sqrt{x^{\prime}}} \ \exp\Bigl(v {(x-x^{\prime}-vt)^2
\over 4\chi x^{\prime}}\Bigr) \ ,
\end{equation}
where $T_{\infty}$ is the temperature as $x \rightarrow \infty$.
Note that this is a function only of $x-vt$, so that in the wall
frame one has a steady distribution.    $\chi$ is
of order a mean free path, $\ell$.  Thus, if $v$ is small,
the temperature varies on a scale of order $\ell /v$,
i.e. on a scale large compared to a mean free path, $\ell$.

Thus, it seems plausible that a steady state situation is achieved.
However, L. Susskind has given a simple example in which this is not
the case, which we refer to as a `snowplow.'
Suppose one has, in the original, unbroken phase, a non-zero density
of some exactly conserved quantum number; we will refer to
these particles as `molecules.' Suppose also, that the molecules
have a big potential energy in the broken phase. Then it is easy to
convince oneself that there is no steady solution; necessarily,
there must be a build up of
molecules as time evolves.  Indeed, one might guess that in this
case there is a layer of molecules in front of the wall, which
grows in size linearly with time.  Moreover, viewed in the frame
of the wall, there is a steady `wind' of particles.  These particles
collide with the ever growing layer of molecules, which is stationary
in the frame of the wall.  The molecules in this wind, presumably,
equilibrate, so the wind comes to rest in the wall frame.  This
buildup of particles leads to an increase
in pressure near the wall of order $\rho v^2$, where $\rho$ is
the molecule density and $v$ the velocity.  Balancing the stresses on
the wall then gives
\begin{equation}\label{41}
v^2 \sim {\delta p \over \rho} \ .
\end{equation}
Note that it is important, in determining the force on the wall, that
the properties of the gas near the wall depend on the velocity of the
wall.

In the Standard Model, there is no exactly conserved quantum number
of this type.  However, there are several approximately conserved
quantum numbers, and one can ask whether something like
this snowplow effect can occur.  Again to develop some intuition,
it is helpful to consider a simple model.
Imagine a wall passing through a region of `sticky dust.'
The dust particles, when struck by the wall, stick to it,
but after they hit the wall they have a lifetime $\tau$.
Then the number of particles per unit area on the wall, $N$,
satisfies
\begin{equation}\label{42}
{dN \over dt} = n_0 v - {1 \over \tau} N \ ,
\end{equation}
where $n_0$ is the density of particles and $v$ is the wall
velocity.  For a steady state, $N= n_0  v \tau$.
In our case, the analog of the dust particles are top quarks,
and the lifetime $\tau$ is some characteristic time to change
the number of tops and antitops (we will discuss this in more detail
below).  Thus, we do in fact expect some velocity-dependent
enhancement of the top quark density near the wall, but
we do not expect that the layer out front should grow indefinitely.
For the rest of this paper, we will assume that the wall does indeed
achieve a steady state.  On the other hand, as we will discuss
shortly,
the enhancement of the density we have discussed above
may be an important factor determining the wall velocity.

Turok  \cite{Turok} has argued that reflection of particles from
the wall does not slow the wall's motion.  His argument starts
with the (correct) observation that
if the velocity and temperature of the gas are constant across
the wall, and if everywhere the system is described by
equilibrium distributions appropriate to the local value of
$\phi$, the force on the wall is independent of the velocity.
It is instructive to consider this part of the argument, and indeed
a modified version of the method for estimating the force on the
wall will be useful in our analysis.

In the wall frame, the net force on the gas is simply
\begin{equation}\label{43}
{dP^i \over dt} = \int d^3 x \ {d^3 k \over 2k^0 (2 \pi^3)}
\, n(\vk, \vx) \, {\pa m^2 \over \pa x} \ .
\end{equation}
With the assumption that the velocity and temperature of the gas
are nearly uniform across the wall, the density in the wall frame
is simply the Lorentz-boosted distribution from the gas frame.  On
the
other hand, $n(\vk, \vx)$ transforms as a Lorentz scalar.  Since
${d^3 k \over 2k^0 (2 \pi^3)}$ is the Lorentz-invariant volume
element, a simple change of variables gives for the net force a
result
independent of the wall velocity.  From this Turok
concludes that one needs to find other sources of dissipation
if the wall is not to accelerate continuously.

Our earlier discussion suggests, however, that the particle
distributions near the wall will exhibit a more complex
dependence on the wall velocity. In other words, even for small
velocity, there will be departures from equilibrium proportional
to the velocity. In the following sections, we will try to take this
velocity-dependence
into account, and estimate the damping of the wall's motion
due to scattering of particles from the wall.  There will, of course,
be other sources of damping; in some regimes, these may be larger.
However, we will see that from this source alone, damping
is sufficient to give non-relativistic motion of the wall for a
wide range of parameters.

The calculations which follow make use of the
infrared improved effective potential described in Section 4 for
which, as already pointed out, baryogenesis has been ruled out
experimentally. However, the study of
the propagation of the bubble wall in a hot plasma
receives its main motivation from the understanding of
the baryogenesis at the weak scale. We will
consequently use the above potential as a toy model in a range
of parameters so chosen that baryogenesis is a viable phenomenon.
To this end, we will often use in what follows the value
$m_{H} = 35$ GeV for which $\phi/T \ \gapp \ 1$.

\subsection{Three Limiting Cases}

This discussion suggests that there are three limiting
cases which one might wish to consider.  In the first,
the wall is thin compared to mean free paths for all scattering
processes, both elastic and inelastic.  Then a typical particle,
as it passes through the wall, loses momentum chiefly through
its interaction with the wall.  Near the wall, one expects
significant, $v$-dependent departures from equilibrium.
A second extreme will occur if the mean free paths for
both elastic and inelastic processes are short compared to
the size of the wall.  In that case, the system will be quite
close to equilibrium; deviations from equilibrium will be
proportional to the velocity and some power of a typical
mean free path.  Finally -- and this is the situation which
we will see is closest to reality -- mean free paths for elastic
processes may be short, but for inelastic processes long.
In this case, in addition to the small deviations from equilibrium
we have just mentioned, there will also be a density enhancement
as in our snowplow discussion.  We expect that the wall will be
slowest if the `thin wall' picture is valid, fastest in the
`very thick wall' case (where mean free paths for inelastic
processes are short) and will have some intermediate velocity
in the third case.

In this and the following two sections, we will consider these
three cases.  After estimating the relevant length scales,
we will make a series of (admittedly preliminary) computations
of the velocity and wall thickness.  We will see that all three
limits suggest a non-relativistic value of the wall velocity.

Before considering the properties of the bubble wall, it is
useful to consider the system at the critical temperature, $T_c$.
At this temperature\footnote{We note that the analysis of Ref.
 \cite{Turok} is indeed valid {\em at} $T=T_c$.}
it should be possible for the phases with broken
and unbroken gauge symmetry to coexist.  Regions of different
phases should be separated by a wall, which we refer to as a bubble
wall or domain wall, at rest.  It is easy to determine the form
of this bubble wall.  One is looking for a static solution of the
equations of motion in the presence of the gas; the relevant
equation is simply
\beq\label{44}
{\partial^2 \phi \over \partial x^2}  = {\partial {V(\phi,T)} \over
\partial \phi} \ .
\eeq
We require that $\phi$ tend to a constant as $x \rightarrow
\pm \infty$, so we can immediately solve the equation by quadrature:
\beq\label{45}
\Delta x = \int{d\phi\over \sqrt{2V(\phi,T)}} \ .
\eeq
For a Higgs mass of $35$ GeV, and a top quark mass
of $120$ GeV, one finds that the wall thickness is
\beq\label{46}
\delta\sim 40\; T^{-1} \ .
\eeq
This estimate of the wall thickness will provide a useful
benchmark in what follows.

Indeed, it is useful to compare this number with the mean free
paths for various processes.  In considering the properties
of the bubble wall, the relevant mean free paths are those
for particles which interact with the wall, i.e. principally
top quarks, $W$'s and $Z$'s. The processes with the shortest
mean free paths are elastic scatterings.  These exhibit the
characteristic singularities of Coulomb scattering at small
angles.  What actually interests us, however, is the momentum
and energy transfer in these collisions.  This is a problem which
has been extensively studied, and we can borrow the relevant
results.  For a relativistic top quark of energy $E$, one
has  \cite{BraThom}
\begin{equation}\label{47}
{dE \over dx} \sim - {8 \pi \over 3} \alpha_s^2 T^2 \ln(E/T) \ .
\end{equation}
We expect a similar formula to hold for $W$ and $Z$ scattering,
with $\alpha_s$ replaced by $\alpha_W$.  This number is to
be compared with the momentum loss due to interaction with the wall:
\begin{equation}\label{48}
{dp_x \over dx} =  - {1 \over 2p_x} {dm^2 \over dx} \ .
\end{equation}
Assuming that the wall thickness is of order $\delta\sim 40\,
T^{-1}$,
and noting that for the $t$ quark, $\Delta m^2 \sim .3\, T^2$,
while for the $W$ and $Z$, $\Delta m^2 \sim .2\, T^2$,
one sees that the momentum loss per unit length due to scattering
is in both cases
much larger than that due to the wall.

We can understand this
result in an alternative way.  The elastic scattering cross section
diverges at small angles, in empty space.  In the plasma, we expect
that this divergence is cut off.  Examining the expression for the
gluon propagator in Ref.  \cite{Gluon}, we see that for the
Coulomb fields, this cutoff is the Debye mass,
\begin{equation}\label{49}
m_{D}^2 = (2 N_c + N_f )\ {g_s^2 T^2 \over 6} \approx  \ 2.5 T^2 \  ,
\end{equation}
while for transverse gauge bosons, it is $m_{D}^2 k_0^2 /3 k^2$.
Actually, the  fact that $m_{D}$ is comparable with  $T$
means that one should go back and self-consistently calculate
the propagators (more or less as we did in our discussion of
the Higgs potential earlier).  We will not do this here,
but instead simply assume that both the transverse and longitudinal
gauge boson exchanges are cut off at a scale of order $T$.
So to get a crude estimate of the mean free path, we simply calculate
the elastic scattering cross section with a gauge boson propagator
\begin{equation}\label{50}
D_{\mn}= {g_{\mn} \over q^2 - T^2} \ .
\end{equation}
One then obtains a total cross section
\begin{equation}\label{51}
\sigma_t \simeq {16 \pi \alpha_s^2 \over T^2} \ .
\end{equation}
To estimate the mean free path, we multiply this by the flux of
quarks. Similarly one can estimate the contribution due to scattering
from gluons, and the mean free paths of $W$'s and $Z$'s.
Finally one obtains an estimate for the
mean free path of order $\ell \sim 4 \, T^{-1}$ for
quarks, and $\ell \sim 12 \, T^{-1}$ for $W$'s and $Z$'s. We will see
below that the wall velocity goes as $\sqrt{\ell}$; we expect that
the uncertainties here will not qualitatively affect those
calculations. It would be
desirable to redo this analysis including the screening
self-consistently and using real transport equations.
However, it is reassuring that they
are consistent with the expression above for the stopping power.

We will also need an estimate of the mean free path for processes
which change the number of top quarks or the number of $W$'s
and $Z$'s. A good measure of this distance is provided by
the lifetimes of these particles in the high temperature plasma.
If we treat the scalar field as approximately constant in space,
then we expect that these decay rates have essentially the same
form as the zero temperature rates,
with the zero-temperature masses replaced by
$\phi$-dependent masses,
and including the
appropriate factor for time dilation. Hence the ratio of widths to
masses is the same as at zero temperature, weighted by the Lorentz
factor. These give numbers of order $1 \%$ or smaller and
thus the mean free path for decays will be of order $100$'s of
$T^{-1}$'s.  While we have not
attempted to do a complete analysis, it appears from an examination
of many cross sections that processes which equilibrate
the various types of particle numbers are not likely to be effective
at distances of order $\delta$.

\subsection{Thin Wall}

The first limiting case we will consider, in some
ways conceptually the simplest, is what we refer to as the
`thin wall.' In this limit, the thickness of the wall, $\delta$,
is less than a typical mean free path $\ell$ for
relevant particles.\footnote{This should not be confused
with the thin wall approximation during the bubble formation,
when the thickness of the wall is assumed to be small as
compared with the radius of the bubble.}  In particular, the
momentum transferred by scattering
to top quarks, $W$'s and $Z$'s (the particles which gain mass
in the broken phase) as they pass through the wall is
small compared to the change in their mass. These particles
constitute roughly $20\%$ of the plasma. The other $80\%$, made up of
light particles, are irrelevant in the sense that they interact only
very weakly with the wall, but play an active role in establishing
a steady state. In this limit, one can compute the force on the wall
semiclassically by assuming some distribution
on either side, and then simply following trajectories of individual
particles across the wall. In the rest frame of the
wall, energy is conserved, so it is easy to compute the momentum
transferred by individual particles to the wall as they pass through
or are reflected back.
To estimate the force on the wall in this case, we will
take advantage of the one dimensional geometry of the problem,
and suppose that the wall is moving in the $ + x$ direction
with velocity $v$.  We will assume that the wall is surrounded by
a plasma of roughly constant temperature and velocity.
  Indeed, our discussion of the diffusion
equation (\ref{40})
suggests that the temperature and other quantities should vary
on a scale $\ell /v$.  For small wall velocity, this
is
much larger than the thickness of the wall. This assumption will be
justified {\em a posteriori}.

We noted in the first section of this paper
that a simple estimate,
eq. (\ref{1}),  was obtained in Ref.  \cite{[3]}.  This estimate is
valid for strongly first order phase transitions, when the main
difference in energy density $\rho$ inside and outside the
bubble is due to heavy particles, which acquire mass $m \gg T$
inside the bubble. Such particles, coming from the phase $\phi = 0$,
 are completely reflected by the
 bubble wall. However, in  the minimal electroweak theory, the
masses of the particles are not very large compared to the
temperature and
one cannot neglect the effect of particles crossing the wall. We have
presented full details of the calculation in the Appendix. We quote
the results here, obtained by expanding in small $v$:
\beq
v
\ce + {\cal O} (v^2) =  V(0,T) - V(m,T) \ .
\label{eq:veloo}
\eeq
Here
\beq\label{eq:coeff}
{\cal E} =  \rho(0, T) - \rho(m, T)
- {m^2 \over 4\pi^2} \int^{\infty}_m  E \, n_o(E)\, {dE}  \ .
\eeq
In the limit of large masses ($m \gg T$), the last  term
in this equation vanishes
and the first two accounts for the energy density contrast
$\Delta\rho(0,T)$, in agreement with (\ref{1}). To study the
properties of
eqs. (\ref{eq:veloo}), (\ref{eq:coeff}) further, we make an
expansion in powers of $m/T$. After a little algebra, we obtain
\beq
\ce= {3 \over \pi} E \,T \phi^3 +
{1 \over 4}\phi^4 \left[ \la-\la_T+2 B
\Bigl( 4\ln{\phi\over v_o} -7\Bigr) \right]
\label{eq:vitfull}\eeq
and
\beq
V(0,T) - V(m,T) = -  \left({T\over T_c} - {\phi_c\over\phi}
(1-\eps)
\right) {\la_{T_c}\over 8} \phi_c \phi^3 \ ,
\label{eq:poteps}\eeq
with $\epsilon$ given by
\beq\label{paraexp}
\epsilon = {T_c^2- T^2 \over T^2_c - T_o^2} \simeq {T_c- T \over T_c
- T_o} \ {}.
\eeq
In the case of small $\eps$,
when $T$ is close to $T_c$ and $\phi\sim\phi_c
(1+\eps)$,  eq. (\ref{eq:veloo}) gives
\beq
v\simeq \frac{\pi}{6}\frac{\eps}{1+\eps}
\left( 1 + \frac{\pi}{6}(1+\eps) \left[ \frac{\la-\la_T}{\la_T}
+ \frac{2B}{\la_T}\Bigl(4\ln\frac{\phi}{v_o} - 7\Bigr) \right]
\right)^{-1} \ .
\label{eq:appr}
\eeq
The relevant value of $\epsilon$ corresponds to the time when the
biggest bubbles propagate in the plasma, which is also the time when
bubbles fill up the universe.  Fig. 6 shows that
$\epsilon$
varies between $0.1$ and $0.3$ for a large range of Higgs
masses and is rather insensitive to the top mass; we use the
value $0.26$ corresponding to a Higgs mass of $35$ GeV
to illustrate our results. For a zero
top mass,\footnote{This is clearly an unphysical value. However, this
information will be useful in the discussion of thick walls in the
next
subsection, since for thick walls
 the top quark contribution is relatively suppressed.}
the term in parenthesis is approximately $1$; in such a case,
\beq\label{oo5}
v\sim {\pi\over 6}{\eps\over 1+ \eps} \sim 0.1 \ .
\eeq
As the top mass increases to a value $> m_Z$, its contribution
becomes dominant and the velocity decreases significantly;
as an example, for $m_t = 120$ GeV, $v \sim 0.06$.
We have analyzed eq. (\ref{eq:veloo}) numerically,
Fig. 7  shows the dependence of the velocity on $\epsilon$ while
Fig. 8 illustrates the dependence of the velocity on the top mass.

We now elaborate further on the situation describing the steady
state.
Since particles change their momentum in crossing or bouncing off the
wall, the local thermal distribution is spoiled; its restoration
requires the release of energy-momentum at a certain rate,
implemented by the flow of light quarks.
Clearly, to satisfy the conservation
of energy and momentum, the plasma has to adjust
its velocity and temperature distributions. For this purpose, the
wall and the non-thermalized particles can be viewed as a source of
energy-momentum of a size of a few mean free paths,
moving in a
relativistic plasma with a non-relativistic velocity. This situation
is similar to the diffusion problem described in Section 6.1.
Qualitatively, one expects a steady state situation in the rest frame
of the source, a rather uniform distribution of temperature and
velocity behind the source extending across the source on a typical
distance $l_D \sim {\ell \over v} $.
{}From our estimate of the
velocity
above, $l_D$ is typically $20 $ times larger than $\ell$ which
justifies
the assumption that the wall is interacting with a plasma of uniform
velocity and temperature. Furthermore, as a good approximation, we
can assume this velocity and temperature to be the ones far away from
the wall. We can then write equations for the conservation of energy
and momentum across the wall:
\beqra
\gamma'^2 \Bigl( v'^2 \rho(m,T') - V(m,T') \Bigr) -
\gamma^2  \Bigl( v^2 \rho(0,T) - V(0,T) \Bigr) =0 \ ,\nonumber \\
\gamma'^2 v' \Bigl( \rho(m,T') - V(m,T') \Bigr) -
\gamma^2 v \Bigl( \rho(0,T) -V(0,T) \Bigr) =0 \ ,
\label{eq:solution}
\eeqra
where $v'=v+\delta v$ and $T'=T+\delta T$ are the quantities defined
behind the wall.

It is easy to compute the changes in velocity $\delta v$
and temperature $\delta T$ of the plasma as it crosses the wall.
These equations and the steady state assumption, together with
equation
\beq
\Bigl( V(0,T') - V(m,T') \Bigr) -v'  \ce' \simeq 0 \ ,
\label{eq:veloo2}
\eeq
yield a unique answer for these quantities.
Typically, ${\delta v \over v} \sim +1\% $, so the velocity, $v$,
is still given accurately by Fig. 7.  For the temperature
variation, one finds
$ {\delta T
\over T}$ is negative and of the order of or less than $ -0.0001\%
$.\footnote{The gradient of temperature, ${\delta T \over T}$,
goes like ${\delta\gamma\over
\gamma} \sim -v\delta v$.}  Furthermore, $\delta v$ is
positive; the process is a deflagration. It is also easy to check
that a small amount of entropy is produced in the plasma, in
agreement
with the second law of thermodynamics.

One can attempt to obtain the shape of the wall by writing eqns.
(\ref{eq:solution}), (\ref{eq:veloo2})
in a differential form.
At $\epsilon=0$, the resulting equation is just that which we studied
earlier for the domain wall separating the coexisting phases.
At finite $\epsilon$, however, there are velocity-dependent
corrections to the equation, and it is more difficult to analyze.
Because the temperature and velocity
vary in space, the problem one has to solve is
analogous to the motion of a particle in a time-dependent potential.
Moreover, the two minima of this potential are not degenerate,
so one needs to understand how the oscillations of the scalar
field in the true minimum are damped.
However, for small $\epsilon$ (and $v$), we do not expect the
shape of the wall to be too much different than for $\epsilon=0$; in
particular, the size of the wall (see eq. (\ref{45}))
is still given to a good approximation by
\beq\label{size}
\delta\sim 2\, {\sqrt{2 \lambda_T} \over E} \sim 40 \ T^{-1}  \ .
\eeq

As this is a several times bigger than the mean free paths of the
relevant
particles ($\ell_{W,Z}\sim 12\, T^{-1}$ and $\ell_{top} \sim
4 \, T^{-1}$),
the thin wall approximation
just described is questionable within the
minimal version of the
electroweak theory. However, this approximation
may be valid in some extension
of the standard model. For instance,
the authors of Ref.  \cite{[5]} have argued that baryogenesis at the
weak scale is viable in a model with a singlet scalar and one Higgs
doublet. The presence of a light singlet makes the transition more
strongly first order,
and, consequently, allows for a Higgs mass above the
experimental bound. In such models, the wall tends to be thinner.

So far, we have assumed for our thin wall analysis that the
density to the left and right of the wall (of particles moving
to the right and left, respectively) are precisely the
equilibrium densities.  However, our discussion earlier
of the snowplow problem suggests that there may be other
effects we should consider as well.   Suppose that there were
only elastic processes, i.e. no processes which changed
the separate numbers of different particles and antiparticles.
Suppose, also, that (for example) all top quarks were reflected
from the wall.  Then we would be in precisely the snowplow
situation, except now in a relativistic version.

More realistically, consider the possibility that, say for
top quarks, the probability of reflection from the wall
of a given quark is $f$.
Suppose, also, that
the mean free path for processes which change the number
of tops or anti-tops is $\tau$.  Then there will
be some buildup of top quarks near the wall.
The time required for
the wall to catch up with a given top quark goes as
$\ell \over v^2$, where $l$ is a typical mean free path.
As a result, in the limit of very small $v$, the wall doesn't catch
up with a typical particle before it undergoes an ``identity
change."  Thus, we have something like the ``sticky dust"
picture described earlier.  Per unit area of the wall,
there is an increase in the number of top quarks
of order $\tau f v n$, where $n$ is the equilibrium top quark
density.
These quarks are spread over a distance of order $\sqrt{\ell\tau}$,
so their density is of order $f v n {\sqrt{\tau\over\ell}}$.
This gives an extra contribution to the force on the wall of order
$v \Delta \rho  n f \sqrt{\tau\over\ell}$, where
$\Delta \rho$ is the free energy difference on the two sides
of the wall.  It is easy to compute $f$; for bosons,
$f = { \zeta(2) \over 2 \zeta(3)}{m \over T}$; for fermions,
the result is $2/3$ as large.
Assuming that the square root in this expression
is a number of order 5-10,
this is comparable to the force we have computed
above, and will tend to decrease the velocity of the wall.
This discussion, of course, is extremely crude, but
it suggests that there are various effects at least as large
as those we have considered above, all of which slow the wall.
Thus, if the thin wall approximation is a good guide, the wall
is likely to be quite non-relativistic.  We will see that there
are similar effects in the the case of a thicker wall.

\subsection{Thick Wall}

We now consider the case that the wall is extremely thick.
As is clear from our earlier discussion of mean free paths,
this, like the thin wall case, is not completely realistic.
These calculations, however, should bracket the true situation.
At the end of this section we will try to estimate the effects
of the density enhancement which occurs because particle numbers
are approximately conserved. So we first consider the case where
the mean free paths for both elastic and inelastic processes are
short.
To get an idea of how finite elastic scattering lengths affect the
velocity of the bubble wall, we assume that particles propagate
freely over distances of order a mean free path, $\ell$.
We view the bubble wall as a succession of slices with thickness of
order $\ell$, and for each of these we repeat
the thin wall analysis.  We refer the
reader to the Appendix for the details of the derivation and
the precise formulae, and just summarize the results here.
We write the result in the following form
\beq
{\cal E} = {\cal S}_b {\cal E}_b^{thin} + {\cal S}_f {\cal
E}_f^{thin} \ ,
\eeq
where ${\cal S}$ are suppression factors dependent on $\ell$.
They are not very sensitive to $m_t$ and $m_H$, and
behave as $(\ell/\delta)^{1/2}$ (see Fig. 9).
Using the values quoted above ($\ell_{W,Z} \sim 12\, T^{-1}$,
$\ell_{top}\sim 4\, T^{-1}$ and $\delta \sim 40\, T^{-1} $), we find
${\cal S}_b \sim 0.45$ and ${\cal S}_f \sim 0.15$.
Using the equations for ${\cal E}$ from the previous section and
assuming $m_H\sim 35$ GeV and $m_t\sim 120$ GeV, we find a velocity
of about
\beq\label{ooze}
v \sim 0.2 \ .
\eeq
Using the values above for ${\cal S}_b$ and ${\cal S}_f$,
Fig. 10 illustrates the velocity for a range of top masses.

In reality, however, we expect, because the numbers of tops and
anti-tops (and similarly $W^{+}$'s and $W^{-}$'s) are not quickly
equilibrated, their densities will be enhanced near the wall;
since this enhancement will depend on the velocity, there will be
a velocity-dependent drag on the wall, similar to that we discussed
above in the thin wall case.

Again, we will content ourselves with an extremely crude estimate.
As in the thin wall case, we imagine that as particles pass through
the wall, some fraction is reflected.  These reflected particles
make a random walk until they decay, or undergo scattering
processes which change their identity, with a lifetime $\tau$.
Then the number of particles per unit area on the wall
is $N= f n v \tau$, where as before, $f$ is the fraction of
particles which are reflected.
This extra density of particles will be spread over the thickness
of the wall, or over $\sqrt{\ell \tau}$, whichever is larger.
For our crude estimate, we will assume this density
is spread uniformly over the wall.  In other words, we will
assume that the density is the equilibrium density (for a given
value of the scalar field), except enhanced by a factor
of the form $1 + {n_0 v \tau \over \ell}$.  In this case,
one obtains an increase force on the wall of order
$\Delta F= {f n v \tau \over \ell}\times \Delta \rho $,
where, as before, $\Delta \rho$ denotes the internal energy
difference on the two sides of the wall.  As an estimate of the
quantity $f$, we take the ratio of the equilibrium densities on
the two sides of the wall.  For top quarks, this gives
a number of order $ 5\%$. The ratio $\tau \over \ell$ is likely
to be of order $5 -10$. Thus, this effect may be even more important
than the effects we have considered above, i.e. we have
probably overestimated the wall velocity.

We can also ask about the wall thickness in this limit.
Our discussion here suggests that the thickness will be modified
from its value at $T_o$ by a factor of the form
$1 + {n_0 v \tau \over \ell}$, i.e. by a small amount.
However, it is somewhat harder to develop a complete
theory of the wall shape in this limit.  For example,
in the analog mechanics problem, if one allows for
spatial variation of the temperature, one has to consider
a system with time-dependent forces.  Moreover, one has
to consider how the `motion' of $\phi$ damps out
in the region of broken symmetry.  This may require consideration
of types of damping other than those we have considered up to
now.  We will not explore this issue further here, and
simply assume that the shape of the wall is only slightly
modified from its form at $T_c$.

\section{Conclusions}

The study of the electroweak phase transition began two decades
 ago, and stimulated work in many areas of what has
come to be called astroparticle physics.  However, until now it was
not very
important to know any details of the theory of this phase transition.
For most
applications it was quite sufficient to know that in the early
universe at a
temperature higher than  about $10^2$ GeV the symmetry between weak
and
electromagnetic interactions was restored. Recently it has become
clear
that if we
wish to investigate the possibility of electroweak baryogenesis, we
must have a
complete and detailed picture of the phase transition,  from the
accurate
computation of the critical temperature to the investigation of the
motion of
the bubble walls. In this paper, we have taken  some steps towards a
systematic
investigation of all relevant features of the first order phase
transitions in electroweak theory.

We have seen how to organize the perturbation expansion
in light of the infrared problems which exist at high temperature.
We have shown that no dangerous linear terms arise in the
effective potential; on the other hand, we have
seen that the coefficient of the cubic term, which is responsible
for the first order nature of the transition, is reduced
by a factor $2/3$ from its lowest order value.  This means, in
particular,
that theories with a single Higgs doublet cannot be responsible
for the observed baryon asymmetry, given the present experimental
limits on the Higgs mass.

We have also understood some
aspects of the strongly first order phase transition relevant to
baryogenesis.
For such theories, we have seen that the phase transition
typically proceeds through the formation of critical bubbles with
thick walls.
We have developed a method of analytic investigation of the
probability of
bubble formation,  which is valid for a large class of theories. With
the help
of the stochastic approach to tunneling, we have found  that
subcritical bubbles are only likely to be important (in the minimal
standard model) for Higgs
mass larger than about $100$ GeV, when the phase transition is second
order or
very weakly first order.

We have considered the problem of
propagation of the bubble wall.
To this end, we have considered the minimal Higgs model
with a light Higgs ($m_H < 35$  GeV).  While this possibility
is ruled out, we believe this theory is a good toy model,
whose features mimic those of more realistic theories
in which baryogenesis is possible.
Investigation of the bubble wall motion turns out to be surprisingly
difficult; indeed, we have identified important mechanisms for
slowing
the wall which have been omitted from previous treatments.
Our own treatment is crude, and may omit additional damping
effects, but it suggests that the bubble wall is typically
non-relativistic or only mildly relativistic in these
theories, and tends to be rather thick.  It would
be interesting to construct theories where this is not
the case, which might realize the scenario of ref.  \cite{11}
in which baryons are produced quite efficiently.

While in this paper we have seen that theories with a single
Higgs doublet cannot give rise to the observed asymmetry,
there is still much work to be done to determine whether
or not baryogenesis can occur in extensions of the minimal
theory.  Needless to say, such extensions are interesting in their
own right, and also because they can provide larger CP  violation
than exists in the minimal model.
Apart from the issues of the phase
transition discussed here, further work on B-violation
rates and the detailed mechanisms of baryon production
is still necessary.  Hopefully the observations contained
in this paper will represent a positive step on the road
to a complete understanding.
\vskip 1cm

{\large \bf Acknowledgements}

  We are grateful to many of our
colleagues for sharing their insights into these problems.
In particular, we thank Lenny Susskind for discussions of
the wall propagation, and suggesting we consider the
``snowplow" problem, and Renata Kallosh for explaining to
us some features of gauge fixing
in spontaneously broken theories.  We also thank
Larry McLerran, Neil Turok and Marcelo Gleiser
for discussing their work with us.

 \vfill
\eject

\section{Appendix:  Calculations of the Wall Velocity}
\renewcommand{\theequation}{A.\arabic{equation}}
\setcounter{equation}{0}

In this appendix we give the details of our calculation of the force
exerted
on the advancing bubble wall by the plasma. As discussed in the text,
we will
assume that there exists a steady state; that is, there is a
well-defined
rest frame of the wall, at sufficiently late times after the
appearance of the bubble. In this frame, we have a (time-independent)
velocity and temperature distribution for each component of the
plasma
that reduce to those of the surrounding Universe at spatial infinity.
We focus on those species that are relatively heavy in the broken
phase;
these will obviously give the greatest contribution.\footnote{The
light
degrees of freedom can not be ignored however, as they are
instrumental in the
approach to and establishment of local kinetic equilibrium.}

In principle to solve this problem in all generality, we would write
Boltzmann equations for each component of the plasma, as well as an
equation of motion for the scalar fields. However, in the simplest
cases,
we can learn a great deal by looking at the equations for local
energy-momentum conservation. These may be written in the form
\beq
\pa_\mu T^{\mn}_{\phi} + \pa_\mu T^{\mn}_{{\rm gas}} = 0 \ ,
\label{eq:emc}
\eeq
where we have (arbitrarily) separated off the classical
(zero-temperature)
stress tensor of the scalar field. We can also write
\beq
-F^{(i)}_x = \pa_x T^{xx}_{(i)} \ ,
\label{eq:force}
\eeq
where $(i)$ labels components of the plasma and
$F^{(i)}_{x}$ is
the force density on the wall of the $i^{{\rm th}}$ species.
Combining
(\ref{eq:emc})
and (\ref{eq:force}), we find the full equation for the scalar field:
\beq
\pa_\mu T^{\mu x}_{\phi} = \sum_i F^{(i)}_x \ .
\label{eq:eom}
\eeq
We now turn to the evaluation of $F_x$.

Consider a volume element of width $s$ in the $x$-direction. For our
analysis, we will assume that particles of the plasma typically
traverse
this distance without interacting; we can calculate the force on this
slice of the bubble wall by following individual particles of the
distribution.  We will present the analysis for a general $s$ below,
but note now that the size of $s$ distinguishes the three cases
considered in the text. Namely, the thin wall scenario corresponds
to $s\sim\delta$, where $\delta$ is the wall thickness, the very
thick
wall corresponds to $s$ infinitesimal, and the intermediate scenario
to
$s\sim\ell$, with $\ell$ some relevant mean free path. We will
suppose
that particles entering the volume element on either side
are described by equilibrium distributions
appropriate to the masses on either side
($m_o$ or $m_1$, with $m_1 > m_o$).  Note that particles leaving
the volume element will not, in general, be thermally distributed;
we assume that equilibrium is restored in a distance of order a mean
free path.
In other words, for particles entering the volume, we are assuming
the distribution
\beq
n(E) = n_o \Bigl(\gamma_v (E-vp_x),T,\mu\Bigr) \ .
\label{eq:n}
\eeq
with $E=\sqrt{\vec p^2 + m_{o,1}^2}$.

We compute the force per unit area as follows. In time $dt$, there
are
$n(p,T) \frac{d^3 p}{(2\pi)^3} |v_x| dt$ particles per unit area
providing a force $\Delta p_x / dt$. Thus, the pressure is
\beq
dP = \int \frac{d^3 p}{(2\pi)^3} n(p,T) (\Delta p_x) |v_x| \ .
\label{eq:pres}
\eeq
We write the integration measure as $dE\;E\;dp_x / 4\pi^2$. There are
various regions of integration, corresponding to whether the
particles have sufficient momentum ($p_x^2>(m^2_1-m^2_o)$)
to go through the slice, or are reflected.

For particles bouncing off the wall, incident from the right, the
momentum transfer is just $-2p_x$. We arrive at
\beq
\Delta P_I = \frac{2}{4\pi^2} \int dE\; dp_x\;
p_x^2 n_o\Bigl(\gamma_v (E-vp_x),T\Bigr)  \ ,
\label{eq:p1} \eeq
where the region of integration is
$[(m_o,m_1)\times (-\sqrt{(E^2-m^2_o)},0)]
+[(m_1,\infty )\times (-\sigma,0)]$ with $\sigma=\sqrt{m^2_1-m^2_o}$.

For particles incident from the right with $|p_x|>\sigma$, we find
$\Delta p_x = -\sqrt{p_x^2 - \sigma^2} - p_x$ and the pressure is
\beq
\Delta P_{II} = \int\frac{dE}{4\pi^2} \; dp_x\; p_x
\left( p_x + \sqrt{p_x^2 - \sigma^2}\right)
n_o\Bigl(\gamma_v (E-vp_x),T\Bigr)  \ , \label{eq:p2}
\eeq
where the region of integration is
$[(m_1,\infty )\times (-\sqrt{(E^2-m^2_o)},-\sigma)]$.

For particles incident from the left, the momentum transfer is
$\Delta p_x = \sqrt{p_x^2 + \sigma^2} - p_x$ and so
\beq
\Delta P_{III} = \frac{1}{4\pi^2} \int dE\; dp_x\; p_x
\left( -p_x + \sqrt{p_x^2 + \sigma^2}\right)
n_o\Bigl(\gamma_v (E-vp_x),T\Bigr) \ , \label{eq:p3}
\eeq
where here the integration region is
$[(m_1,\infty )\times (0,\sqrt{(E^2-m^2_1)})]$.

We need to now integrate over the thickness of the wall. If there are
$N \sim \delta/s$ slices, we write
\beq
F_x = \sum_{n=0}^{N} \Delta P_n \ , \label{eq:f}
\eeq
where here $\Delta P_n$ is the sum of eqs.
(\ref{eq:p1})-(\ref{eq:p3})
with $m_o$ the mass at the $n^{{\rm th}}$ step, given by
\beq
m_o^2 = \int_{+\infty}^{x_n} dx \frac{dm^2}{d\phi}\frac{d\phi}{dx}
\sim m^2 \frac{n^2}{N^2} \ , \label{eq:m}
\eeq
In the last step we have made a linear approximation to the wall
profile.

We may proceed in a variety of ways. If we assume that the speed
of the wall is small, we can expand the distribution (\ref{eq:n})
to lowest order in $v$.  On the other hand, we could expand in
powers of $m/T$, at the risk of making a small error for the top
quark.
In this Appendix, we will choose the former expansion. We find
\beq
n_o\Bigl(\gamma_v (E-vp_x)\Bigr) = n_o(E) - vp_x\;\frac{\pa n_o}{\pa
E}
(E) + O(v^2) \ .
\eeq
To lowest order we have:
\beqra
\Delta P_n^{[0]} & = & \frac{2}{3} \int^{\infty}_{m_o}
\frac{dE}{4\pi^2}\
 n_o(E,T)\, (E^2-m^2_o)^{3/2} - ( m_o \leftrightarrow m_1 ) \nonumber
\\
         & = & F(m_o,T) - F(m_1,T) \ .
\eeqra
Summing over the wall and over species, we find the simple result
\beq
F^{[0]}_x = F(0) - F(m^2) \ .
\eeq
The pressure difference, to
lowest order, is the difference of free energy densities. We have
neglected contributions proportional to gradients in temperature and
chemical potential.

The next-to-leading
order terms account for the forces of the plasma on the wall.
To order $v^1$, we find
\beqra
\Delta P^{[1]}_n & = & -2v\int^{m_1}_{m_0}
\frac{dE}{4\pi^2}\  n_o(E)\, E(E^2-m_o^2) \\ \nonumber
&  & -v\int^{\infty}_{m_1}\frac{dE}{4\pi^2}\  n_o(E)\, E
\left(\sqrt{(E^2-m_1^2)} - \sqrt{(E^2-m_o^2)} \right)^2 \ .
\label{eq:ov}
\eeqra
The next step is to sum this over the wall. We will do this for the
three cases separately. For the thin wall, we simply take $N=1$,
$m_o = 0$ and $m_1 = m$, and obtain
\beq
F^{[1]}_x = v \left (\rho(0) - \rho(m)
- m^2 \int^{\infty}_m \frac{dE}{4\pi^2} E n_o(E) \right ) = v{\cal E}
\ .
 \label{eq:thin}
\eeq
It is  useful to expand this expression in powers of $m/T$.
For $W$ and $Z$ bosons, we find
\beq
\ce^{thin}_{b}\simeq \frac{3}{\pi} ET\phi^3
+ \frac{3\phi^4}{64\pi^2 v_o^4} \left[ 2m_W^4 \left( \ln \frac{m_W^2
\phi^2}{a_B v_o^2 T^2} -{7\over 2}\right) + m_Z^4 \left(
\ln\frac{m_Z^2
\phi^2}
{a_B v_o^2 T^2} -{7\over 2}\right) \right] \ .
\eeq
Likewise, for top quarks, we find\footnote{The expansion is
less trustworthy here; in particular, we have neglected terms of
order $({m \over T})^5$.}
\beq
{\cal E}^{thin}_{f}\simeq -\frac{3 \phi^4}{16 \pi^2 v_o^4} m_t^4
\left[ \ln \frac{m_t^2 \phi^2}{a_F v_o^2 T^2} -{7\over 2} \right]
+ O(({m\over T})^5)
\eeq
The total $\ce$ is then
\beq
\ce \simeq \frac{3}{\pi} ET\phi^3 + {1\over 4} \phi^4
\left[ \la-\la_T + 2B\left( 4\ln {\phi\over v_o} -7 \right) \right] \
{}.
\eeq

If the wall is much thicker than any relevant mean free path, we take
$s$ proportional to $dx$. We can
easily see that (A.14) has no linear term in $dx$, and
we recover Turok's result \cite{Turok} that
there
is no $v$-dependence in the force on the wall in the limit $\ell
\rightarrow 0$.\footnote{We
have explicitly checked this through order $v^2$.} Presumably there
are other effects that are ignored in this computation that slow down
the wall.

The case of greatest interest is finite $\ell/\delta$. We have
performed the analysis numerically and find ${\cal E}$ is suppressed
essentially by a factor of $(\ell/\delta)^{1/2}$. (This non-analytic
behavior is expected of eq. (A.14), since its Taylor
expansion
has singularities at order $\sigma^4$.) We can now write
$\ce$ in the following form:
\beq
\ce = {\cal S}_f \ce^{thin}_f + {\cal S}_b \ce^{thin}_b \ .
\eeq
where
\beq
{\cal S}_i = \frac{\ce_i}{\ce^{thin}_i} \ .
\eeq
${\cal S}_f$  and ${\cal S}_b$ are
rather insensitive to $m_t$ and $m_H$.
We have plotted these factors in Fig. 9 {\em vs.}
$(\ell/\delta)^{1/2}$. The behavior of this plot is adequately
approximated by a linear function:
\beq
{\cal S}_i \sim
\left( \frac{\ell}{\delta}\right)^{1/2}
\eeq
for both fermions and bosons. We find then, that the
wall velocity, for $m_H = 35$ GeV and $m_t = 120$ GeV,
 is approximately $v \sim  0.2$ (see Fig. 10).

\vfill
\eject

\pagebreak

\end{document}